\documentclass[12pt,a4paper]{article}
\usepackage{graphicx}  
\usepackage{dcolumn}   
\usepackage{bm}        
\usepackage{amssymb}   
\usepackage{subfigure}
\usepackage{amsmath,amsfonts}
\usepackage{setspace}
\usepackage[utf8]{inputenc}
\usepackage{authblk}
\begin{document}

\title{Reexamination of the mean-field phase diagram of biaxial nematic
liquid crystals: Insights from Monte Carlo studies}
\author{B. Kamala Latha, Regina Jose, K. P. N. Murthy, and V. S. S. Sastry}

\affil{School of Physics, University of Hyderabad, Hyderabad 500046,
Telangana, India}

\begin{abstract}

Investigations of the phase diagram of biaxial liquid crystal systems 
through analyses of general Hamiltonian models within the simplifications 
of mean-field theory (MFT), as well as by computer simulations based on 
microscopic models, are directed towards an appreciation of the role of 
the underlying molecular-level interactions to facilitate its spontaneous 
condensation into a nematic phase with biaxial symmetry. Continuing 
experimental challenges in realising such a system unambiguously, 
despite encouraging predictions from MFT for example, are requiring 
more versatile simulational methodologies capable of providing insights 
into possible hindering barriers within the system, typically gleaned 
through its free energy dependences on relevant observables as the system 
is driven through the transitions. The recent brief report from this group 
[B. Kamala Latha, \textit{et. al.}, Phys. Rev. E \textbf{89}, 050501(R), 2014], 
summarising the outcome of detailed Monte Carlo simulations carried out 
employing entropic sampling technique, suggested a qualitative modification 
of the MFT phase diagram as the Hamiltonian is asymptotically driven towards 
the so-called partly-repulsive regions. It was argued that the degree of 
the (cross) coupling between the uniaxial and biaxial tensor components of 
neighbouring molecules plays a crucial role in facilitating, or otherwise, 
a ready condensation of the biaxial phase, suggesting that this could be a 
plausible factor in explaining the experimental difficulties. In this 
paper, we elaborate this point further, providing additional evidences 
from curious variations of free-energy profiles with respect to the 
relevant orientational order parameters, at different temperatures 
bracketing the phase transitions. 
\end{abstract}

\maketitle

\section{Introduction}
                                  
The thermotropic biaxial nematic phase which was predicted by Freiser 
\cite{freiser}  nearly four decades ago has attracted considerable 
attention recently for various reasons, ranging from a fundamental 
question of conducive experimental conditions for its realization to
its envisaged applications in display devices. Though predictions made by  
various mean-field (MF) theoretic treatments \cite{Straley} - \cite{matteis07},
Landau free energy based analyses \cite{Alben} - \cite{Mukherjee09} and 
computer simulations \cite{luck80} - \cite{preeti11} support the feasibility
of such a phase, success on the experimental front has been rather modest 
\cite{luck04}. Experimentally the biaxial phase was first obtained in a 
lyotropic, ternary mixture of potassium laurate, 1-Decanol and $D_{2}O$ in 1980
\cite{Yu} and more recently in bent-core compounds \cite{Acharya04, madsen}, 
organo-siloxane tetrapodes \cite{Merkel, Figueirinhas05}, 
LC polymers \cite{Severing} and colloidal systems of Goethite particles 
\cite{vandenpol09}. Though recent experiments \cite{Date,Kouwer} point to 
low transition enthalpies for rod-disc systems, an unambiguous biaxial 
phase has not been established in such systems. From the point of view 
of application, it is anticipated that  the minor director could switch 
more readily compared to the major director in an external field 
\cite{Lee07, BerardiA} leading to faster  response times. Even in the case 
of recent bent-core molecules there appears to be a 
debate on the consistency in the experimental findings 
\cite{VanLe, Mamatha,Vaupotic, Ostapenko}. Achieving spontaneous  
macroscopic biaxiality in nematic liquid crystal phases with appreciable 
biaxial order appears at the moment to be a challenge. 

   The recent theoretical studies on the other hand point to a 
more optimistic picture: they predict that the condensation of a biaxial 
phase could occur over a wide range of the Hamiltonian parameter space of a 
general quadrupolar model \cite{Sonnet} - \cite{matteis07}. However, 
the analysis of the mean field model was noted to be unsatisfactory, as 
the phase behaviour of the biaxial system in the limit of vanishing 
intermolecular biaxial interaction traversing in the process the 
so-called partly repulsive region of the Hamiltonian was found to be 
contravening the biaxial phase stability criterion \cite{matteis07}. In 
this context we revisit the  mean-field phase diagram with detailed 
Monte Carlo simulations.  Main results of this study were 
briefly presented recently \cite{kamala14}. The other MC work on the 
so-called $\mu$-model \cite{Dematteis08} was also similarly concerned
with the consequences of the contribution of a repulsive interaction 
term in the  Hamiltonian.  
 
     In this paper, we present the details of a qualitatively different
type of Monte Carlo sampling that we adopted for the study. It was 
observed that the sampling methods to extract equilibrium averages 
based on equilibrium ensembles (constructed using the Metropolis 
algorithm \cite{Metropolis}) largely lead to results in accord with the 
MFT in the so-called  attractive region of the Hamiltonian parameter space. 
Keeping this in  view, we adopted the Wang-Landau sampling procedure \cite{wang}
augmented by frontier sampling \cite{Zhou, jayasri09} to determine the 
representative  density of states of the system, enabling the 
calculation of all relevant thermodynamic properties.
We  find that this more versatile and efficient technique 
results in qualitatively different results in certain regions of the
 parameter space, leading to the proposal of a modified  phase diagram
(relative to MFT). We argue that, such differences, which develop 
progressively as the 'partly repulsive region' is reached, are important
in understanding the relative roles of different contributions to the 
intermolecular tensor interactions.
 
   The mean field Hamiltonian model employed and its representation 
 for purposes of simulation are outlined in section II. The sampling 
 technique and the simulation details are presented in section III. The
 observations based on these computations are presented in section IV, 
 followed by conclusions in section V.
       
\section{Hamiltonian model}
The MF analysis \cite{Sonnet} - \cite{matteis07}, 
is based on the general quadrupolar orientational Hamiltonian, proposed by
 Straley \cite{Straley} and set in terms of tensors
\cite{Sonnet}. Accordingly, the interacting  biaxial molecules
are represented by two pairs of symmetric, traceless tensors 
($\bm{q}$, $\bm{b}$) and ($\bm{q^{'}}$, $\bm{b^{'}}$). Here $\bm{q}$ and 
$\bm{q^{'}}$ are uniaxial components about the unit molecular vectors $\bm{m}$ 
and $\bm{m^{'}}$, whereas $\bm{b}$ and $\bm{b^{'}}$ (orthogonal to
$\bm{q}$ and $\bm{q^{'}}$, respectively), are biaxial. These irreducible
components of the anisotropic parts of susceptibility tensor are 
represented in its eigen frame 
$(\bm{e},\bm{e_{\perp}},\bm{m})$ as
 
\begin{subequations}\label{s1}
\begin{align}
\bm{q} &:= \bm{m} \otimes \bm{m} - \frac{\bm{I}}{3} \\
\bm{b} &:= \bm{e} \otimes \bm{e} - \bm{e}_{\perp} \otimes \bm{e}_{\perp}
\end{align}
\end{subequations}
where $\bm{I}$ is the identity tensor. Similar representations hold for 
$\bm{q^{'}}$ and $\bm{b^{'}}$ in the eigen frame
 $(\bm{e^{'}}, \bm{e^{'}_{\perp}}, \bm{m^{'}})$. The interaction energy 
is written as

\begin{equation}
H=-U[\xi \, \bm{q} \cdot \bm{q}^{\, \prime}
+ \gamma(\bm{q} \cdot \bm{b}^{\, \prime} + \bm{q^}{\, \prime} \cdot \bm{b}) +
\lambda \, \bm{b} \cdot \bm{b}^{\, \prime}]
\label{eqn:w2}
\end{equation}
where $U$ is the scale of energy, $\xi$ = $\pm 1$, $ \gamma$ and
$\lambda$ are dimensionless interaction parameters, determining the 
relative importance of the uniaxial-biaxial coupling and 
biaxial-biaxial coupling interactions between the molecules, respectively.

\begin{figure}
\centering
\includegraphics[scale=0.4]{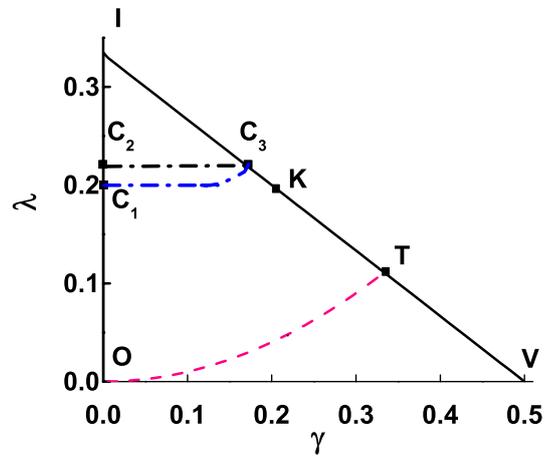}
\caption{(color online) Essential triangle : Region of biaxial stability. 
OI and IV are uniaxial torque lines intersecting at the point I ($0,1/3$). 
OT is the dispersion parabola which meets the line IV at the Landau point T. 
Point V $(1/2, 0)$ is the limit of biaxial stability for the interaction. 
$C_{1}$ $(0, 0.2)$ and $C_{3}$ $(5/29, 19/87)$ are tricritical points and 
$C_{2}$ $(0.22,0)$ is a triple point (G. De Matteis $et  \ al$, \textit
{Continuum Mech. Thermodyn.} \textbf{19}, 1-23 (2007)). K $(0.2,0.2)$ is 
a point where $\mu$ = -1 (refer to text).}
\label{fig:1}
\end{figure}
       
\indent   Mean-field analysis of the Hamiltonian identifies a triangular 
region  OIV in the $(\gamma,\lambda)$ plane - called the essential triangle - 
representing the domain of stability into which any physical system
represented by Eqn.~\eqref{eqn:w2} can be mapped \cite{Bisi06, matteis07}
(see Fig.~ \ref{fig:1}). The dispersion parabola $\lambda=\gamma^{2}$
 \cite{Luck75} traverses through the interior of the triangle, intersecting
 IV at the point T, called the Landau point. Region of the triangle above 
 the parabola corresponds to a Hamiltonian where all the terms are 
 attractive, while the region below is noted to be partly repulsive 
 \cite{Bisi06}. In particular, a mean-field (MF) phase diagram was 
predicted \cite{matteis07} as a 
function of the arc length OIV (Fig.~ \ref{fig:1}), denoted by 
$\lambda^{*}$, defined as  $\lambda^{*}=\lambda$  on the segment OI, and
$$ 
\lambda^{*}=\dfrac{({1+\sqrt{13}\gamma})}{3},
$$ 
with
 $$
\gamma =\dfrac{({1-3\lambda)}}{2}
$$
covering the segment IV. The MF phase diagram predicts for  
$\lambda^{*}\lesssim 0.22$ ($\gamma =0, \lambda \lesssim 0.22)$  a two stage
transition from the isotropic to a biaxial phase, with an intervening uniaxial
nematic phase. The uniaxial-biaxial transition is computed to be second 
order $(N_{B}=N_{U}-I)$ upto the point 
$C_{1}$ $(\gamma = 0,\lambda \simeq 0.2)$, and then changes 
to first order $(N_{B}-N_{U}-I)$, till $C_{2}$ 
$(\gamma = 0, \lambda \simeq 0.22)$.  For the 
rest of the range of $\lambda^{*}$, a direct isotropic-biaxial transition
 is expected, extending upto V in Fig.~\ref{fig:1}. This transition is 
 predicted to be first order $(N_{B}-I)$ for $\lambda^{*}\leq 0.54 \ $ 
$ (C_{3}, \lambda^{*} = 0.54,(\gamma= 5/29, \lambda = 19/87))$ and 
 second order $(N_{B}= I)$  upto the point V $(\gamma = 0.5, \lambda = 0.0)$.
Hence $C_{1}$ and $C_{3}$ are tricritical points and $C_{2}$ 
is a triple point.
 
       We consider here the diagonal form of the interaction Hamiltonian 
in Eqn.~\eqref{eqn:w2} \cite{Bisi06, matteis07} expanded as a superposition 
of two quadratic terms, i.e
$$             
H=-U(a^{+}\bm{q}^{+} \cdot \bm{q}^{+'}+a^{-}\bm{q}^{-} \cdot \bm{q}^{-'})
$$
where $\bm{q}^{+}$ and $\bm{q}^{-}$  are orthogonal molecular biaxial 
tensors represented as
$$
\bm{q}^{\pm}= \bm{q}+\gamma^{\pm}\bm{b}
$$
with
$$
\gamma^{\pm}=\dfrac{3\lambda-1\pm\sqrt{(3\lambda-1)^{2}+12\gamma^{2}}}{6\gamma}
$$
$$
a^{+}=\dfrac{\gamma^{-}-\gamma}{\gamma^{-}-\gamma^{+}}
$$
and
$$
a^{-}=\dfrac{\gamma-\gamma^{+}}{\gamma^{-}-\gamma^{+}}.
$$
Along OI where $\gamma=0$, $\bm{q}^{+}=\bm{q}$, $ \bm{q}^{-}=\bm{b}$, 
$a^{+}=1$ and $a^{-}= \lambda$ implying that $\bm{q^{+}}$ is pure uniaxial 
and $\bm{q^{-}}$ is pure biaxial and the Hamiltonian reduces to an 
interaction in terms of a single parameter $\lambda$

\begin{equation}
H= -U( \bm{q} \cdot \bm{q}^{\prime }+ \lambda \bm{b} \cdot  \bm{b}^{\prime}).
\label{eqn:w3}
\end{equation}

Similarly, along IV, defined by $1 - 3\lambda -2 \gamma = 0$, the Hamiltonian
is expressed in terms of uniaxial tensor $\bm{q}_{2}^{*}$ and biaxial tensor
$\bm{b}_{2}^{*}$ as \cite{matteis07}

\begin{equation}
H = -U\dfrac{1-\lambda}{2}(-\mu \ \bm{q}_{2}^{*} \cdot \bm{q}_{2}^{* \prime}
    + \bm{b}_{2}^{*} \cdot \bm{b}_{2}^{* \prime})
\label{eqn:w4}
\end{equation}

where
$$
\bm{q}_{2}^{*} = -\dfrac{1}{2}\bm{q}^{-}=(\bm{e} \otimes \bm{e} - \dfrac{\bm{I}}{3})
$$
$$
\bm{b}_{2}^{*} = \dfrac{3}{2}\bm{q}^{+} = (\bm{m} \otimes \bm{m} - \bm{e}_{\perp} \otimes \bm{e}_{\perp})
$$
and 
$$
\mu=\dfrac{(1-9\lambda)}{(1-\lambda)}.
$$
The pair-wise interaction in Eqn.~\eqref{eqn:w4} now reduces to

\begin{equation}
H=U^{'}[\,{\mu(\bm{e}\otimes \bm{e}-\dfrac{\bm{I}}{3})\cdot(\bm{e}^{'}\otimes
\bm{e}^{'}-\dfrac{\bm{I}}{3})-(\bm{e}_{\perp}\otimes
\bm{e}_{\perp}-\bm{m}\otimes \bm{m})\cdot(\bm{e}^{'}_{\perp}\otimes
\bm{e}^{'}_{\perp}-\bm{m}{'}\otimes \bm{m}^{'}})].
\label{eqn:w5}
\end{equation}
with $ U{'}={U(1-\lambda)}/{2}.$ 
 In this format, $\mu=-3$ 
corresponds to the point I (0, 1/3) in  Fig.~\ref{fig:1}, 
$\mu=0$ to the  Landau point T ($1/3,1/9$) (LP), and
$\mu=+1$ to V (0.5, 0.0). In particular, we observe that $\mu= -1$ 
corresponds to $\lambda^{*}\backsimeq 0.57$ located at K $(0.2, 0.2)$ in 
Fig.~\ref{fig:1}.

  For simulation purposes, the general Hamiltonian in Eqn.~\eqref{eqn:w2} is 
 conveniently recast as a  biaxial mesogenic lattice model, where particles 
of $D_{2h}$ symmetry, represented by unit vectors $\bm{u}_{a}$, $\bm{v}_{b}$ 
on lattice sites a and b interact through a nearest-neighbour pair 
potential \cite{romano}

\begin{equation}
U= -\epsilon \lbrace G_{33} - 2\gamma(G_{11}-G_{22})+
\lambda[2(G_{11}+G_{22})-G_{33}]\rbrace.
\label{eqn:w6}
\end{equation}

Here  $f_{ab}$= ($\bm{v}_a$.$\bm{u}_b$),
 $G_{ab}$=$P_2$($f_{ab}$) with $P_{2}$ denoting the second Legendre 
polynomial. The constant $\epsilon$ (set to unity in
simulations) is  a positive quantity setting the reduced temperature
$\textit{T}^{'}=k_{B}\textit{T}/\epsilon $, where $\textit{T}$ is 
the absolute temperature of 
the system. This is recast along IV of the triangle, using Eqn.(17) 
in reference \cite{Dematteis08}, in terms of the parameter $\mu$ as

\begin{equation}
H=\epsilon[\mu G_{11}+(-2G_{33}-2G_{22}+G_{11})].
\label{eqn:w7}
\end{equation}

\section{Details of Simulation}
The Wang-Landau (WL) sampling \cite{wang} is a flat histogram technique 
designed to overcome energy barriers encountered, for example,
 near first order transitions, by facilitating
a uniform random walk along the energy ($\textit{E}$) axis through an 
appropriate algorithmic guidance. The sampling, originally developed for  
Hamiltonian models involving random walks in discrete configurational 
space, continues to be applied to various problems in statistical
physics \cite{ Landau1, Murthy1}, polymer and  protein studies 
\cite{Rathore03, Seaton10, Priya11} and 
is being developed for more robust applications for continuous systems
\cite{Poulain,Sinha09,Raj,Yang13, Vogel13, Katie14, Xie14} and self 
assembly \cite{Landau13}. The proposed algorithm was modified \cite{jayasri} 
to suit lattice models like the Lebwohl-Lasher interaction \cite{LL}, 
allowing for continuous variation of molecular orientations. It was 
subsequently augmented with the so-called $\textit{frontier}$ sampling 
technique \cite{Zhou, jayasri09} to simulate more complex systems like 
the biaxial medium. The WL sampling is based on effecting a convergence 
of an initial distribution over energy $\textit{E}$ to the density of 
states (DoS) $g(E)$ of the system iteratively. Frontier sampling technique
is an algorithmic guidance, provided in addition to the WL routine, by 
which the system is constrained to visit and sample from low entropic 
regions. The modified Wang-Landau algorithm using entropic sampling 
augmented by frontier sampling \cite{jayasri09} is described below. 

We consider a cubic lattice (size: $L \times L \times L, L= 15, 20$) 
with each lattice site representing a biaxial molecule, and hence 
hosting a (right-handed) triad of unit vectors. We initiate the process 
by assigning random orientations of all the axes at every site, and 
compute the energy of the system at the chosen point in the 
$(\gamma, \lambda)$ plane with the Hamiltonian in
Eqn.~\eqref{eqn:w6} (corresponding to  $ \xi =1$ in Eqn.~\eqref{eqn:w2}), 
under periodic boundary conditions, - the temperature is thus  
measured in reduced units.  The energy range of interest of the system 
($E_{min}, E_{max}$) is divided into N bins (we set N = 40 $L^{3}$) of 
equal width, and the bin energies are indexed as $E_{i}$, corresponding 
to the values at the centre of the $i^{th}$ bin. We indexed these bins 
starting from $E_{min}$. We initialise $g(E_{i})$ to an array 
$g^{(0)}(E_{i})$ with equal values $(i=1,...., N)$, where the 
 superscript is the iteration run index and the subscript is the energy 
 bin index. The estimate of $g(E_{i})$ is improved by updating iteratively, 
until it converges to the density of states within a set tolerance limit.

 For liquid crystal systems with continuous degrees of freedom for the 
 random walk in configuration space, we find it necessary to perform 
 the simulations on a log-log scale to avoid issues of large numbers and 
 consequent overflow problems. Following \cite{Berg}, we work with 
 $\zeta_{i}= \log (\alpha_{i} )= \log(\log (g(E_{i})))$, where 
 $\alpha_{i}$ represents the microcanonical entropy. The acceptance 
 criterion as well as  reweighting procedures are implemented on this 
 scale. 
   
   During the random walk, the system is permitted to transit from an
initial configuration with an instantaneous value $\zeta_{i}$ to a
trial configuration with $\zeta_{t}$ with a probability given by
 
 \begin{equation}
p=min\lbrace1,\exp[-\exp[\zeta_{t}+\log(1-\exp(-(\zeta_{t}-\zeta_{c})))]]\rbrace.
\label{eqn:c215}
\end{equation}

 We update the values of $\zeta_{i} \ (i=1,...., N)$ of the bins with a
Gaussian centered at the accepted bin energy value (say $E_{0}$), as

\begin{equation}
\zeta_{i} \rightarrow \zeta_{i}+\gamma _{0}\exp(\dfrac{-(E_{i}-E_{0}}{\delta})^{2}.
\label{eqn:c216}
\end{equation}  
 
Here $(\gamma_{0}, \delta)$ represent the modification parameters. We 
 kept $\delta$ constant through the simulation (at $0.002 *N $) and chose
 the initial value of $\gamma = 0.1$. Random walk of the system over the
 energy bins, at this value of $\gamma_{0}$, is carried out for a large 
 number of lattice sweeps (attempted $L^{3}$ moves), typically $10^{7}$ 
 sweeps or more depending on the system size. The $\gamma_{0}$ value is 
reduced to $\gamma_{0}\rightarrow 0.95 \gamma_{0}$, and the procedure is 
repeated until $\gamma_{0}$ reaches a set small value close to zero 
($~ 10^{-4}$). The computations involving a progressive reduction of 
$\gamma_{0}$, starting from the initial high value, to the set low value 
constitute an iteration. After two such successive  iterations, the 
differences between histogram values at each bin are determined. If the 
differences are nearly uniform over some energy range, it implies that 
this region is adequately sampled. We expect, on entropic grounds, that 
the flatness of the distribution, in terms of fairly uniform increments 
of histogram values, is achieved more readily starting from the maximum 
energy value. We call the limiting lower energy bin, satisfying the flatness
criterion, as the $\textit{frontier}$, say $E_{c}$. Following the 
suggestion of \cite{Zhou}, we update the values of the histogram above $E_{c}$
by a uniform value (say, 0.5). This makes the system, under the above 
acceptance criterion, to perform random walk preferentially in the lower 
energy region hosting less accessible states,- until the histogram values
 build up to match the values in the higher energy regions, above $E_{c}$.  
This process is continued with such iterations, and new frontiers
are identified at progressively lower energy values, corresponding to an 
approximate estimation of the DoS over larger energy ranges, until the 
frontier reaches $E_{min}$. 

Consequently, a long smoothing run 
is performed (no frontiers are identified at this stage) starting  with 
initial values of  $(\gamma_{0}, \delta)$ set to (0.001, $0.002 *N $) 
and the value of $\gamma_{0}$ is progressively decreased during this 
computation until it reaches practically  zero value, $\simeq 10^{-9}$. Such 
iterations continue until a specified flatness criterion is met over 
the entire energy range. This ensures that the final $\zeta(E_{i})$ converges 
to its asymptotic  value and is  representative of the  density of states 
of the system, within the tolerances prescribed by the flatness criterion.

 We  now construct a large entropic ensemble of microstates (say, 
 $M \sim 4 \times 10^{7}$)  by effecting a random  walk of the system 
 over the energy bins ($i=1,..., N$) with an acceptance probability based 
 on $\zeta^{-1}(E_{i})$ (analogous to Eqn.~\eqref{eqn:c215}).
 We label the  microstates as $C_{\nu}^{i} \ [i=1,....N, \nu = 1,.....M$] 
 with $M \gg N $. We note that an $i^{th}$ bin  for example hosts a large 
 number of microstates $(C_{\nu}^{i})$ with distinct energies 
 $E(C_{\nu}^{i})$, however represented by the same density of states 
 $g(E_{i})$. 

The relevant thermodynamic quantities are calculated at each temperature 
by constructing appropriate canonical ensemble of states using a reweighting 
technique \cite{Swensden}. We refer to these ensembles 
as RW-ensembles, to differentiate from those constructed through the 
Metropolis guided random walk (B-ensembles). The equilibrium averages of 
a physical variable 'O' at a temperature $\textit{T}$
($\beta$=$\frac{1}{\textit{k}_\textit{B}\textit{T}}$) are computed through 
this procedure as
  
  \begin{equation}
\langle O\rangle= \frac{\sum_{C_{\nu}^{i}} O(C_{\nu}^{i})g(E_{i}) \exp{[-\beta E(C_{\nu}^{i})]}}{\sum_{C_{\nu}^{i}} g(E_{i}) \exp[-\beta E(C_{\nu}^{i})]}.
\label{eqn:w8}
\end{equation}
 
The representative free energy $\textsl{F}$, as a function of the energy 
of the system, as well as of the two dominant order parameters (uniaxial 
and biaxial orders) is computed from the DoS and the 
microcanonical energy, - both available as a function of bin number
in the entropic ensemble. 

The WL simulations were carried out at different values of 
$(\gamma, \lambda)$ in Eqn.~\eqref{eqn:w6} so 
as to trace the trajectory OIV of the essential triangle in Fig.~\ref{fig:1}
at about 60 chosen points. For purposes of comparison, conventional MC 
sampling (based on Metropolis algorithm) was used to construct canonical 
(Boltzmann) ensembles. Considering an attempted  N= $L^{3}$ moves as one 
lattice sweep (MC step), the system is equilibrated, and a production run 
is carried out, each for $6 \times 10^{5}$ MC steps. In our analysis, we 
find it necessary to distinguish between the averages from  
B-ensembles and RW-ensembles. 

The physical parameters of interest in this system, calculated at 
each $\lambda^{*}$, are the average energy $<E>$, specific heat $<C_{v}>$, 
energy cumulant $V_{4}$ (= $1-<E^{4}>/(3<E^{2}>^{2})$) which is a measure 
of the kurtosis \cite{Binder}, the four order parameters of the phase 
calculated according to \cite{Biscarini95,Robert} and their susceptibilities. 
These are the uniaxial order $<R^{2}_{00}>$ (along the primary director), 
the phase biaxiality $<R^{2}_{20}>$, and the molecular contribution to 
the biaxiality of the medium  $<R^{2}_{22}>$, and the contribution
to uniaxial order from the molecular minor axes $<R^{2}_{02}>$. 
 
 The averages are computed at a temperature resolution of 0.002 units in 
 the temperature range [0.05, 2.05]. The temperature  $\textit{T}^{'}$ 
 of the  simulation is scaled to conform to  the values used in the mean 
 field treatment: ${1}/{\beta^{*}}= {3\textit{T}^{'}}/{(9[2U(1+3\lambda)])}$ 
 \cite{Bisi06, matteis07}. Statistical errors in different observables 
 are estimated over ensembles comprising a minimum of $5 \times 10^{5}$ 
 microstates, and these are compared with several such equilibrium ensembles 
 at the same  $(\gamma,\lambda)$ value, but initiating the random walk 
 from different arbitrary points in the configuration space. We find the 
 relative errors in energies are 1 in $10^{5}$, while those in the estimation 
 of the order parameters are 1 in $10^{4}$. We also note that these error 
 estimates from RW- ensembles are smaller relative to B-ensembles of 
 comparable size by at least an order of magnitude, owing to the efficacy 
 of the  importance sampling involved in the reweighting procedure.

\section{ Results } 

\begin{figure*}
\centering {\subfigure []{
\includegraphics[scale=0.3]{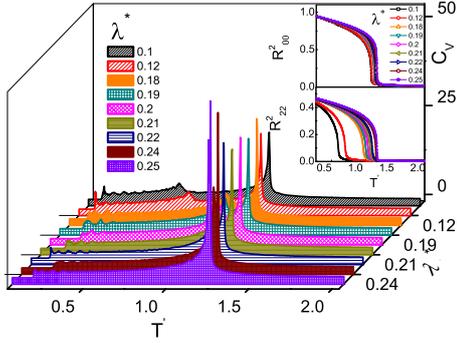}
\label{fig:2a}
}
 \subfigure []{
   \includegraphics[scale=0.3]{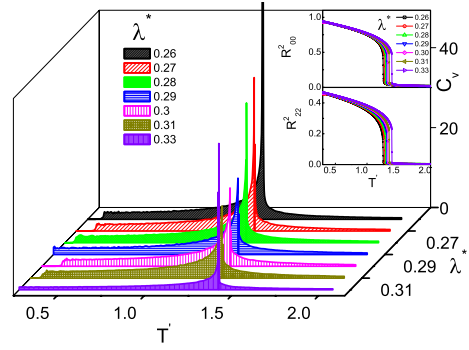}
\label{fig:2b}   
 }
 \subfigure []{
\includegraphics[scale=0.3]{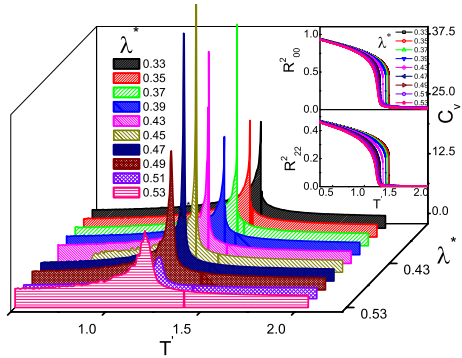}
\label{fig:2c}}
 \subfigure []{
   \includegraphics[scale=0.3]{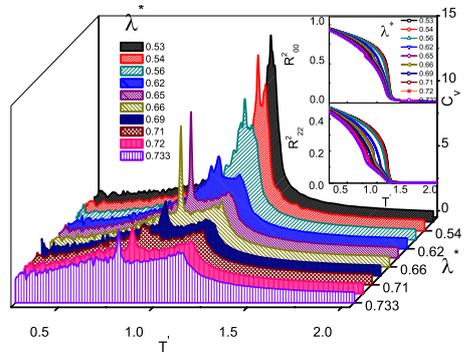}
 \label{fig:2d}
}
}
\caption{(color online) Temperature variation of the specific heat 
(in arbitrary units) for different ranges of $\lambda^{*}$: (a) 0.1 - 0.25; 
(b) 0.26 - 0.33; (c) 0.33 - 0.53; (d) 0.53 - 0.733. Corresponding variations 
of the two  primary order parameters ($R^{2}_{00}$ and $R^{2}_{22}$) are 
shown in the insets with same colour scheme. Splitting of the $C_{v}$ peaks in 
 (d) for $\lambda^{*} > 0.53$ and qualitative changes in the temperature 
 variation of order parameters are clearly observed.  } 
\label{fig:2}
\end{figure*}
    
The temperature variations of the specific heat and the two dominant scalar 
order parameter ($R^{2}_{00}$ and $R^{2}_{22}$) values obtained from 
RW-ensembles at various values of $\lambda^{*}$ along the arc OIT 
($\lambda^{*}$-axis) are shown in Figs.~\ref{fig:2a} - \ref{fig:2d}. 
 
 It is noted from Fig.~\ref{fig:2a} that for all values of $\lambda^{*}$
in the range 0.1 - 0.25, two transition peaks are observed in the specific 
heat. As the biaxial system is cooled from the high temperature isotropic 
phase, an initial $I-N_{U}$ transition occurs at a high temperature $\textit{T}_{1}$ 
followed by a second transition $N_{U}-N_{B}$ at lower temperature $\textit{T}_{2}$. 
The $I-N_{U}$ transition temperature remains  fairly constant with the 
variation in  $\lambda^{*}$, whereas $N_{U}-N_{B}$ transition  shifts 
towards higher temperatures as $\lambda^{*}$ increases from 0.1 to 0.25.  
This behaviour is also reflected in the order parameter profiles shown in 
the inset. The two transitions eventually coalesce at $\lambda^{*}=0.26$ 
resulting in a triple point and a direct isotropic-biaxial ($I-N_{B}$) 
transition occurs from $\lambda^{*}$ = 0.26 to 0.53, as depicted by the 
specific heat profiles and order parameters (inset) of
Figs. ~\ref{fig:2b}- \ref{fig:2c}. These results from RW-ensembles agree
qualitatively with those obtained from the B-ensembles in 
the range of $\lambda^{*} = 0.1-0.53$.

\begin{figure}
\centering {\subfigure []{
\includegraphics[scale=0.25]{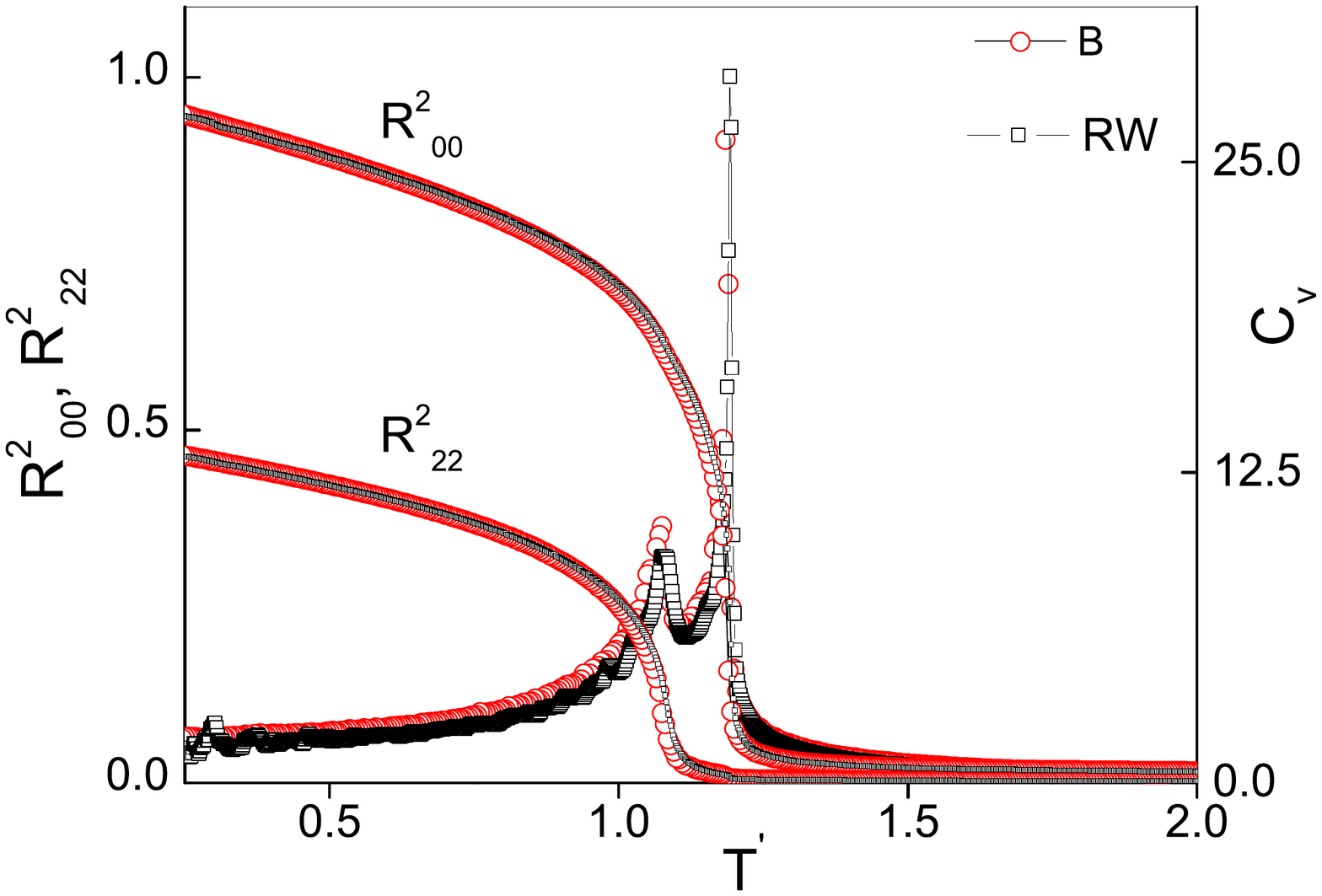}
\label{fig:3a}
}
 \subfigure []{
   \includegraphics[scale=0.25]{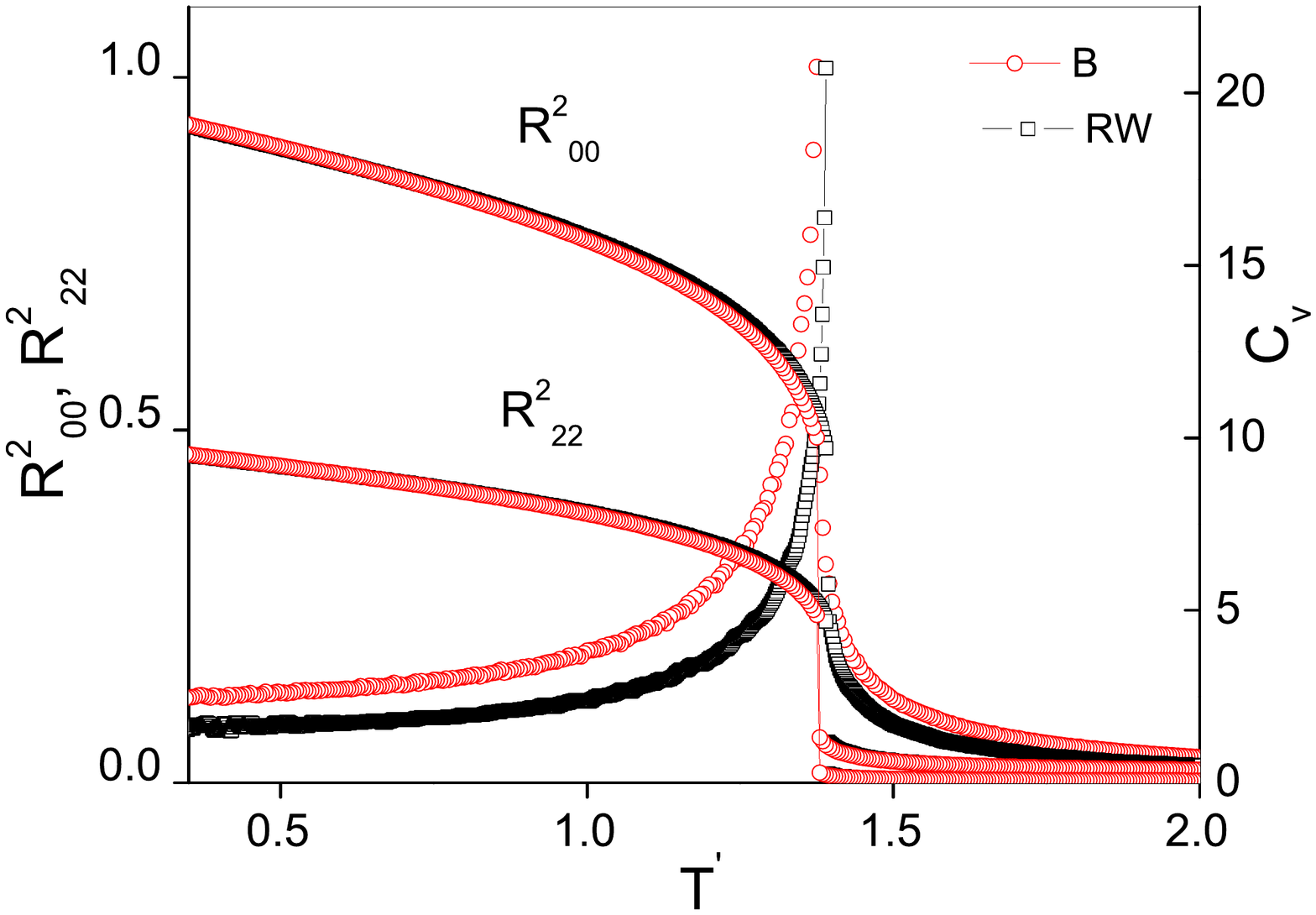}
\label{fig:3b}   
 }
 \subfigure []{
\includegraphics[scale=0.25]{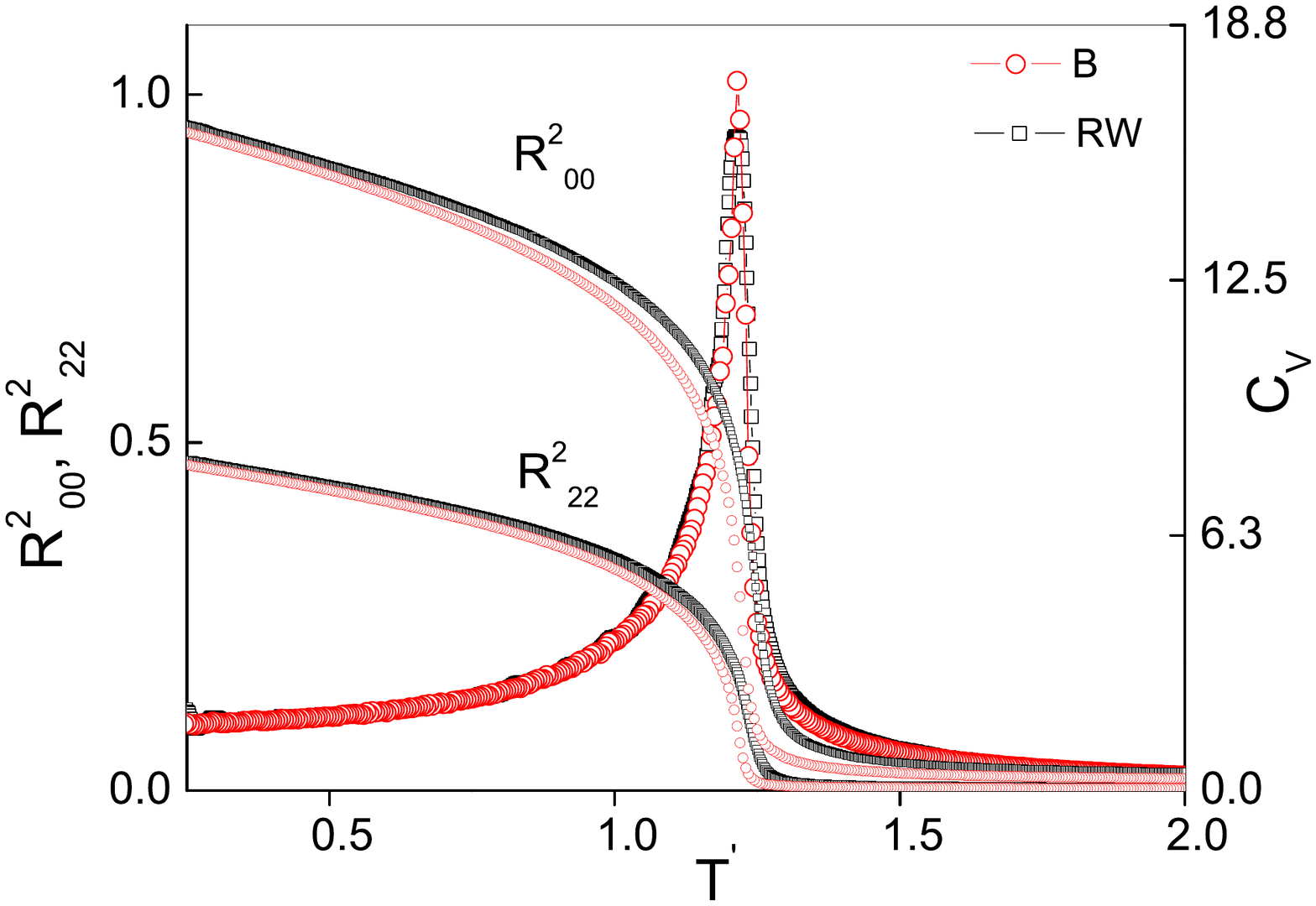}
\label{fig:3c}}
 \subfigure []{
   \includegraphics[scale=0.25]{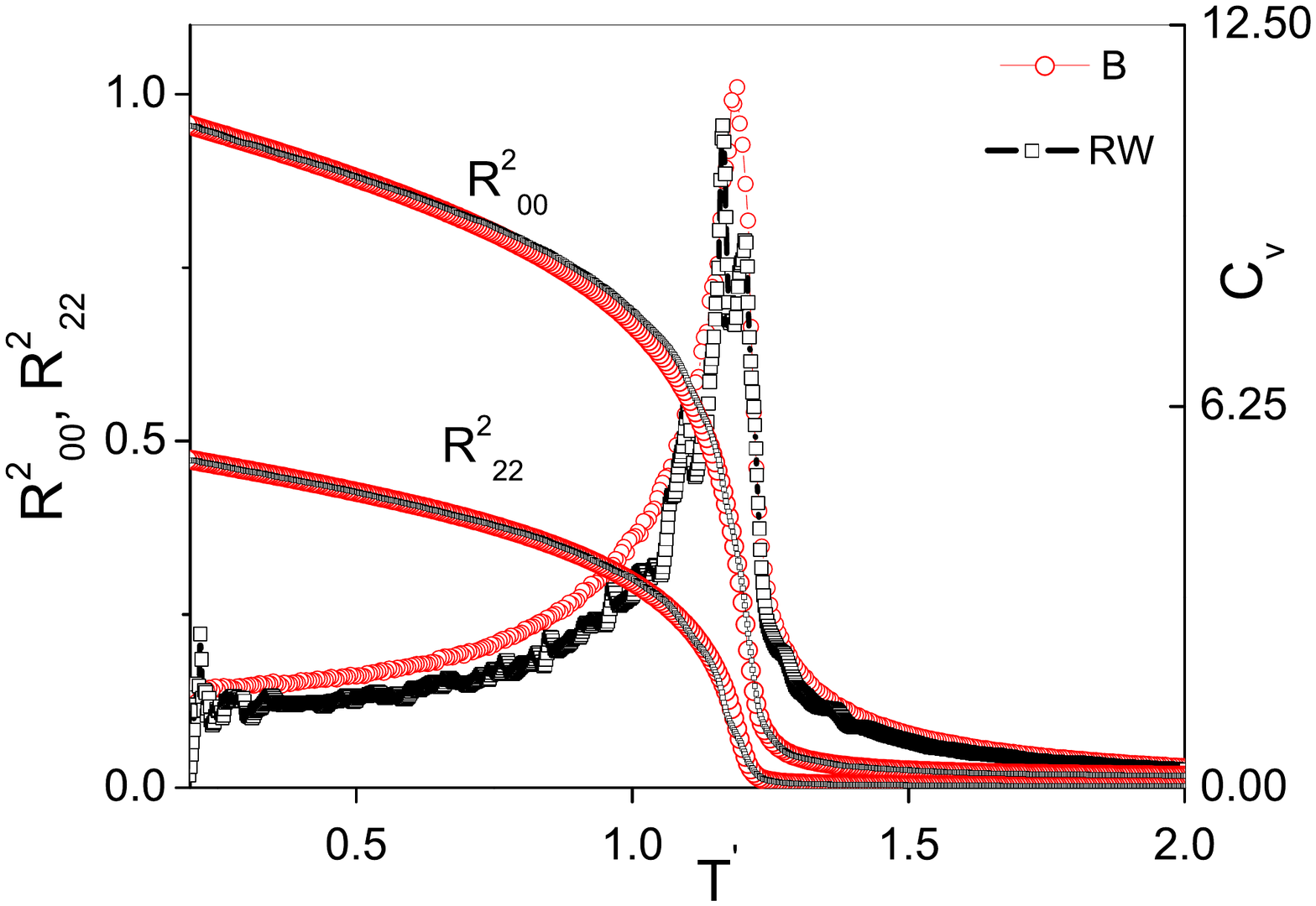}
 \label{fig:3d}
}
}
\caption{(color online) Comparison of the results obtained from B-ensembles 
(hollow red circles) and the RW-ensembles (hollow black squares): Temperature 
variation of the specific heat (in arbitrary units) and corresponding 
variations of the two primary order parameters ($R^{2}_{00}$ and 
$R^{2}_{22}$) are shown for different values of $\lambda^{*}$ in regions 
OI and IV: (a) 0.2; (b) 0.33; (c) 0.51; (d) 0.54. The overlap of the 
corresponding curves (a to c) clearly indicates the agreement between 
the two ensembles upto $\lambda^{*} = 0.53$ as mentioned in the text. 
Fig. ~\ref{fig:3d} shows  the qualitative disagreement first noticed 
at $\lambda^{*}=0.54$.  
} 
\label{fig:3}
\end{figure}

A comparative study of the WL and MC simulation 
results for certain representative values of $\lambda^{*}$ are shown in  
Figs.~\ref{fig:3a} - \ref{fig:3d}. It is observed that qualitative 
agreement with the mean-field predictions exists upto $\lambda^{*}\leq 0.53$
and deviations of the RW-ensembles from MF and B-ensembles start 
from $\lambda^{*}=0.54 (5/29, 19/87)$ (i.e $C_{3}$ in the essential 
triangle of Fig.\ref{fig:1}).

Referring to Fig.~\ref{fig:2a}, the results from  RW-ensembles agree with 
MF predictions except for the actual values of the location of the 
tricritical and triple points $C_{1}$ and $C_{2}$. In this respect, one 
has to make allowances for unavoidable finite size effects on the 
simulation data on the one hand, and the inherent approximate nature of 
 the mean field theoretical analysis in this respect, on the other.

\begin{figure}
\centering 
\includegraphics[scale=0.3]{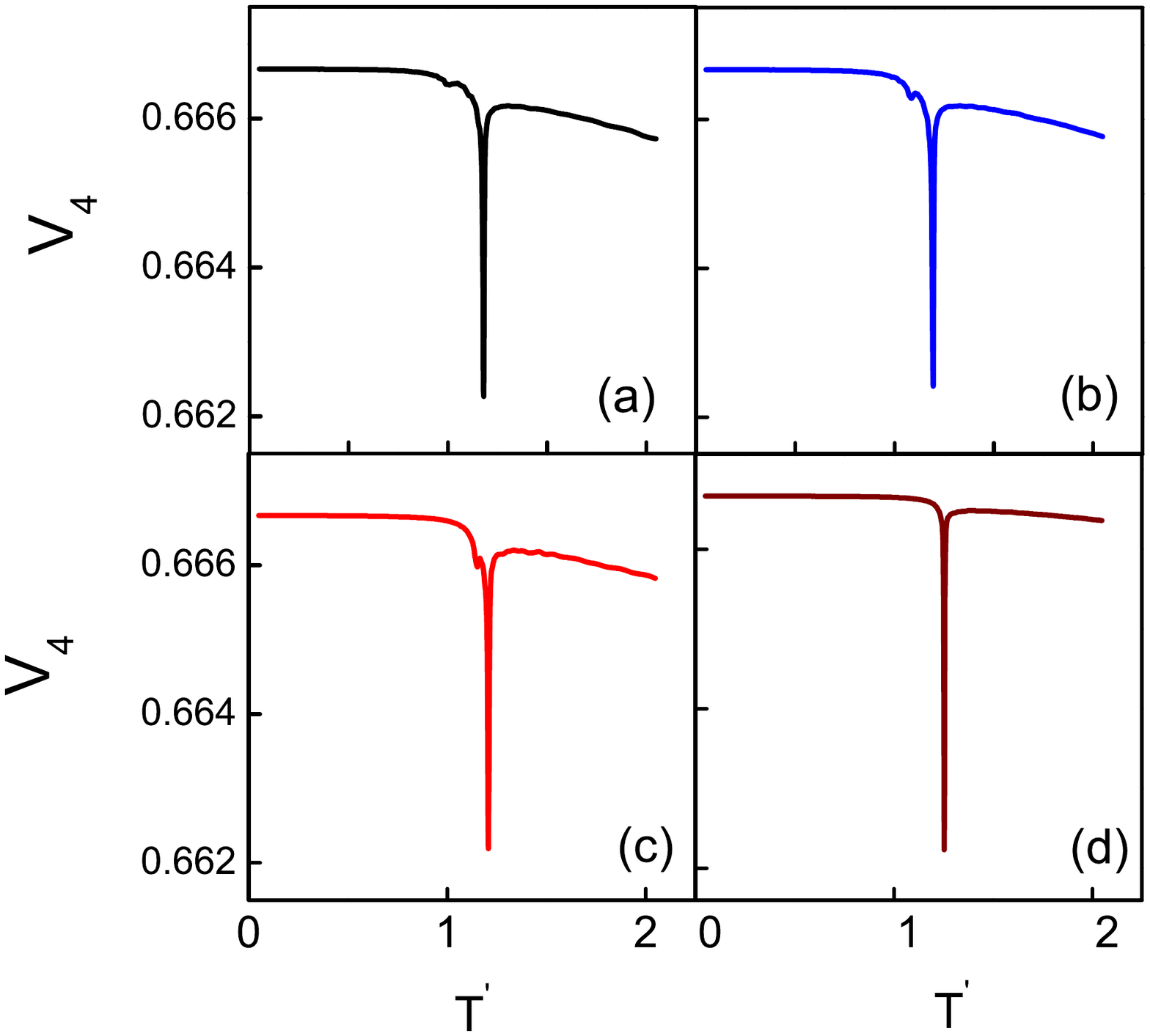}
\caption{ (color online) Variation of energy cumulant with temperature, 
at different $\lambda^{*}$ values:  (a) 0.18; (b) 0.2; (c) 0.22 (d) 0.26} 
\label{fig:4}
\end{figure}   
 
 At L=20, the simulation results show that the tricritical point 
lies in  the neighbourhood of $\lambda^{*}$  = 0.18; the nature 
of the $N_{U}-N_{B}$ transition appears to change to a (weak) first 
order for values of $\lambda^{*}\geq 0.18$, as evidenced from the energy 
cumulant data shown in Fig.~\ref{fig:4}. The triple point is located at 
$\lambda^{*}\sim 0.26$ (corresponding  MF value is $\sim 0.22)$ as the transition 
sequence $I-N_{U}-N_{B}$  changes to $I-N_{B}$ at this value of $\lambda^{*}$ 
(see Fig.~\ref{fig:2b}). Transition temperatures derived from these 
 simulations are summarised in Table \ref{tab:table1}.     

\begin{table}
\caption{ Transition temperatures in the range of $\lambda^{*}$= (0.18, 0.26):
$\textit{T}_{1}^{'}$ and $\textit{T}_{2}^{'}$ are transition temperatures 
(in reduced units) obtained from simulation, while $\textit{T}_{1}^{*}$ and  
$\textit{T}_{2}^{*}$ are the corresponding equivalent mean field temperatures.}
 \begin{center}
  \begin{tabular}{ | p{0.5cm}| p{0.5cm} | p{0.5cm} | p{0.5cm}| p{0.5cm} | }
    \hline
  \centering$\lambda^{*}$&$\textit{T}_{1}^{'}$ & $\textit{T}_{2}^{'}$&$\textit{T}_{1}^{*}$& $\textit{T}_{2}^{*}$ \\ \hline
 \centering 0.18 & 1.1753  & 0.9919  & 0.1272  & 0.1074   \\ \hline
 \centering 0.2  & 1.1937  & 1.0770  & 0.1243  & 0.1122    \\ \hline
 \centering 0.22 & 1.2110  & 1.1490  & 0.1216  & 0.1153     \\ \hline
 \centering 0.26 & 1.2516  &         & 0.1172  &      \\
     \hline
    \end{tabular}
    \label{tab:table1}
\end{center}
\end{table}

\begin{figure}
\centering
\includegraphics[scale=0.3]{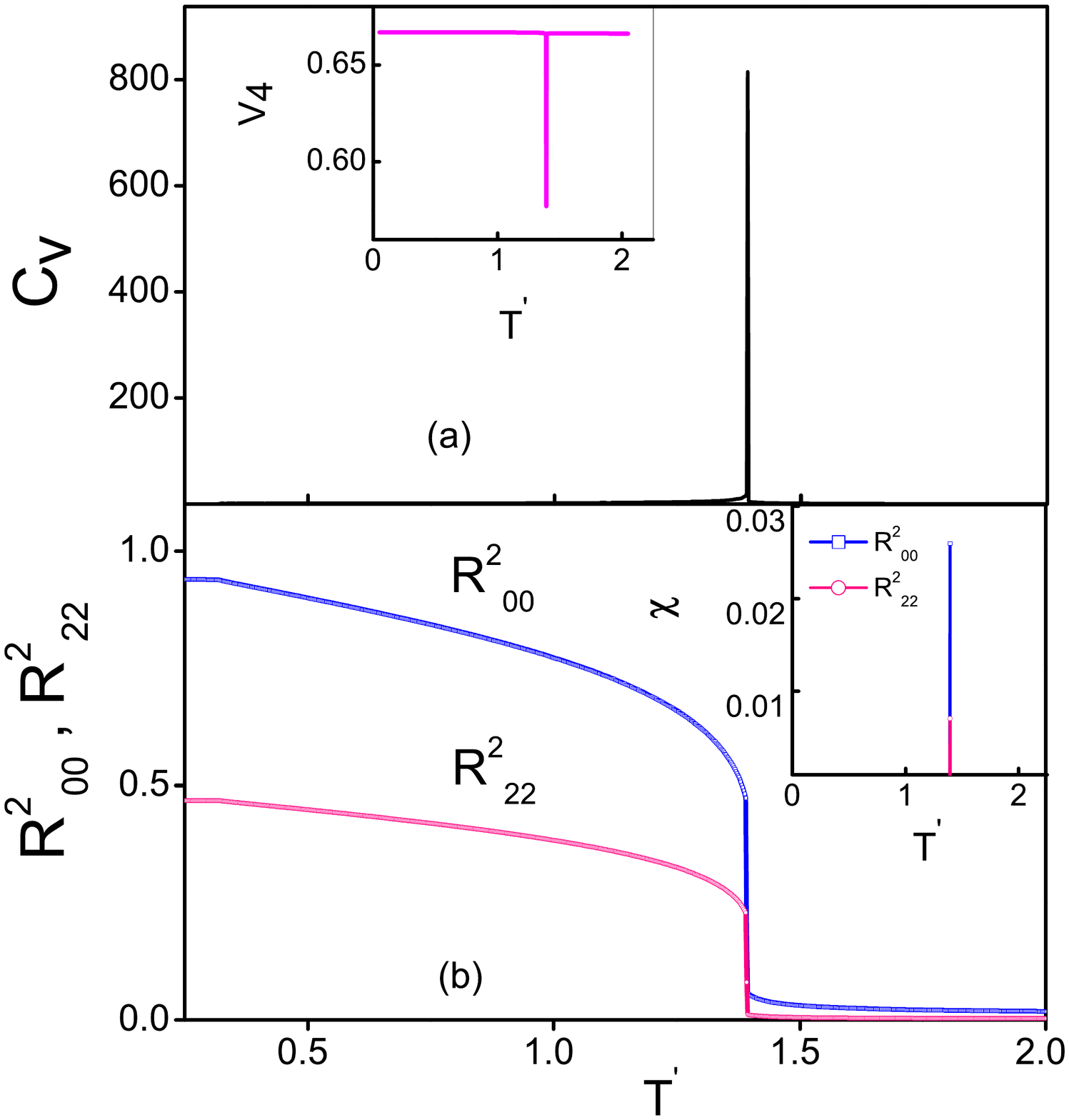}
\caption{(color online) (a) Specific heat profile with (inset) energy cumulant 
(b) Order parameters with (inset) susceptibility profiles for 
 $\lambda^{*}$=0.33. Point I of the essential triangle is the 
 intersection point of the three uniaxial torque axes \cite{matteis07} 
 and hosts the strongest first order $I-N_{B}$ transition, as shown 
 by the very sharp features of these physical properties.}
\label{fig:5}
\end{figure}
           It is of interest to observe that the $I-N_{B}$ transition 
progressively becomes very strong first order as $\lambda^{*}$ value 
increases to $1/3$ and is most pronounced at the point corresponding 
to coordinates $(0,1/3)$ (Fig.~\ref{fig:1}). Fig.~\ref{fig:5} 
depicts the specific heat with (inset) energy cumulant and order parameters
with (inset)  their susceptibilities for this value of $\lambda^{*}$. 
 \begin{figure}
\centering
\includegraphics[scale=0.3]{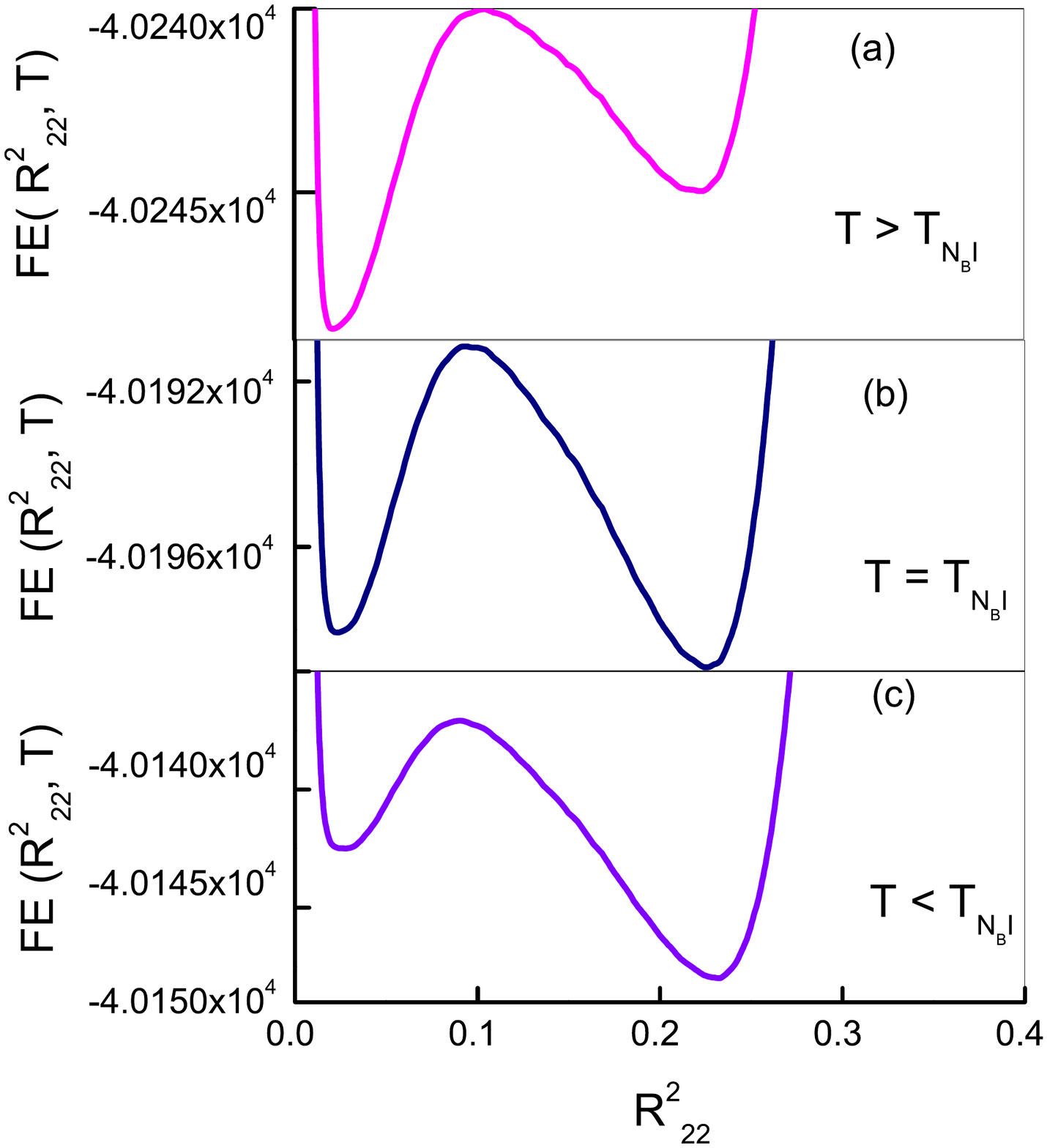}
\caption{(color online) Representative free energy (in arbitrary units) 
as a function of biaxial order parameter $R^{2}_{22}$ at 
(a) $\textit{T} > \textit{T}_{N_{B}I}$ (b) $\textit{T} = \textit{T}_{N_{B}I}$ 
and $\textit{T} < \textit{T}_{N_{B}I}$ at $\lambda^{*}$=0.33 for a 
system of size L=15.} 
\label{fig:6}
\end{figure}
This feature of the transition is also demonstrated by the variation of the 
representative free energy obtained from the DoS and bin energies, as a 
function of the two order parameters. As an example, we depict its 
variations across the transition ($\lambda^{*}$ = 0.330, L=15), in 
Fig.~\ref{fig:6}. Observation of such a strong free energy barrier (see
the coexistence region at $\textit{T}=\textit{T}_{N_{B}I}$) is supportive of the 
MF prediction at I. Very similar plots result as a function of 
$R^{2}_{00}$   also.
  
\begin{figure*}
\centering {\subfigure[]{
\includegraphics[scale=0.22]{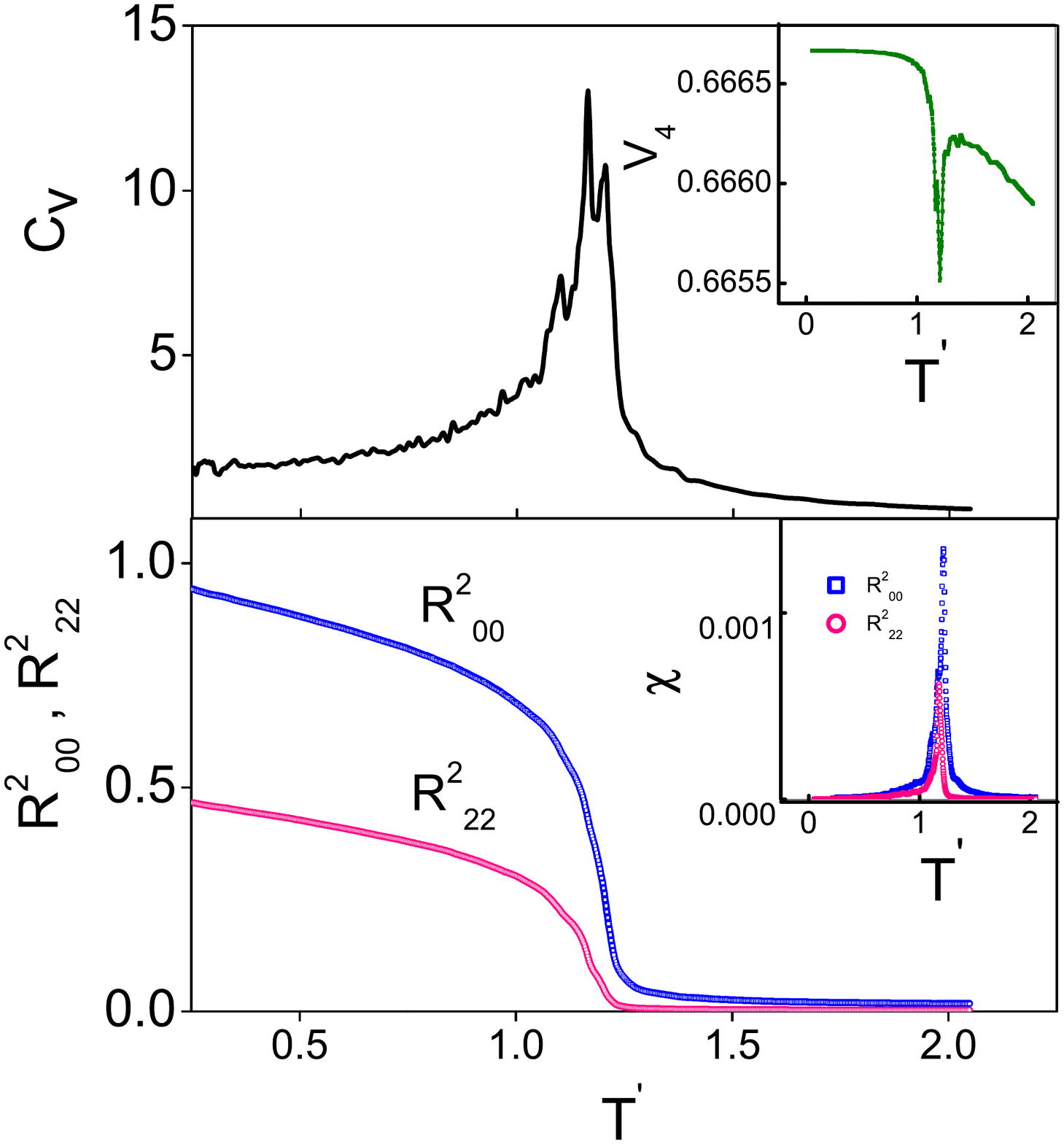}
\label{fig:7a}}
 \subfigure[]{
   \includegraphics[scale=0.22]{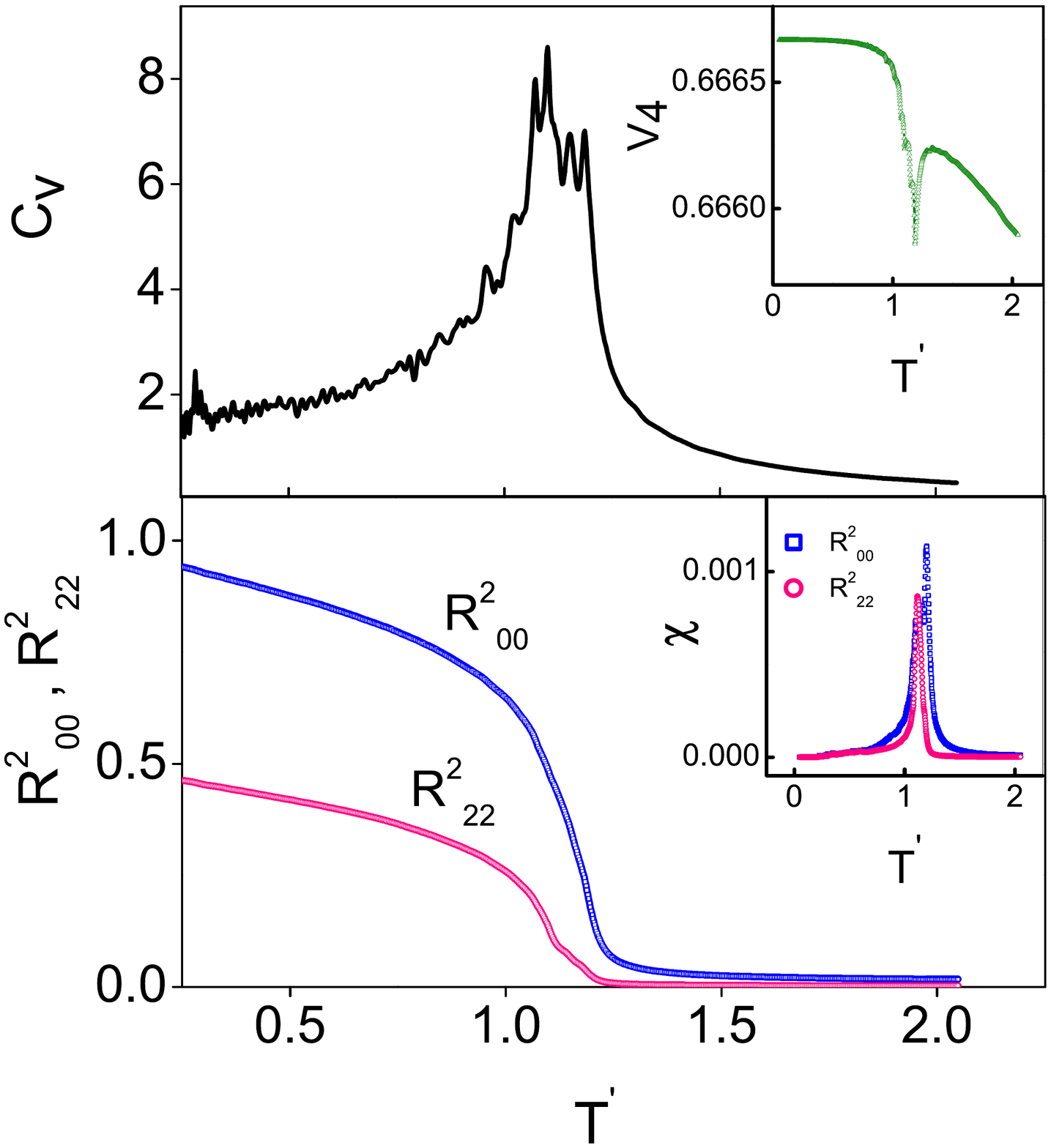}
   \label{fig:7b}}
\subfigure[]{
   \includegraphics[scale=0.22]{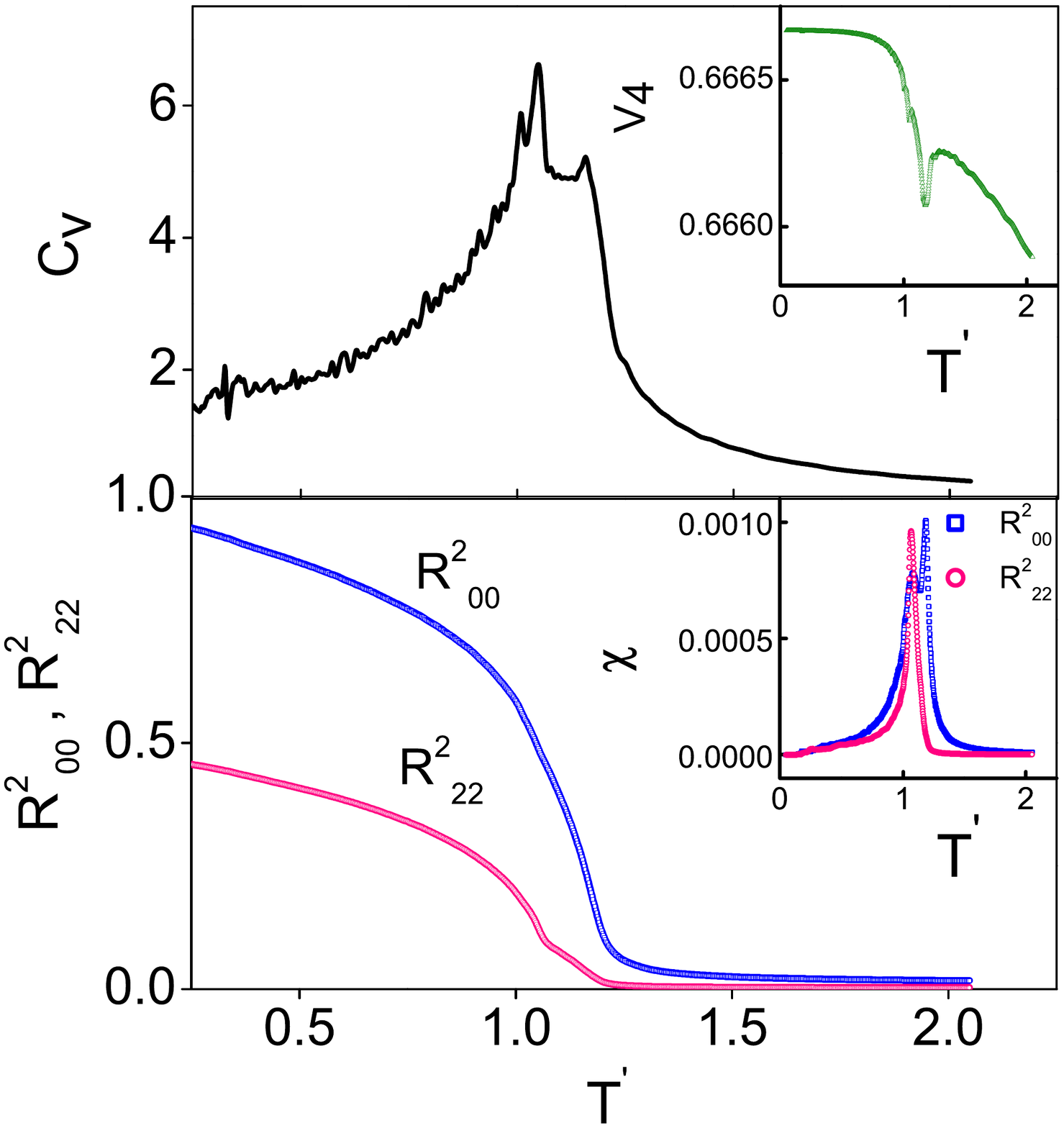} 
\label{fig:7c}}
\subfigure[]{
   \includegraphics[scale=0.22]{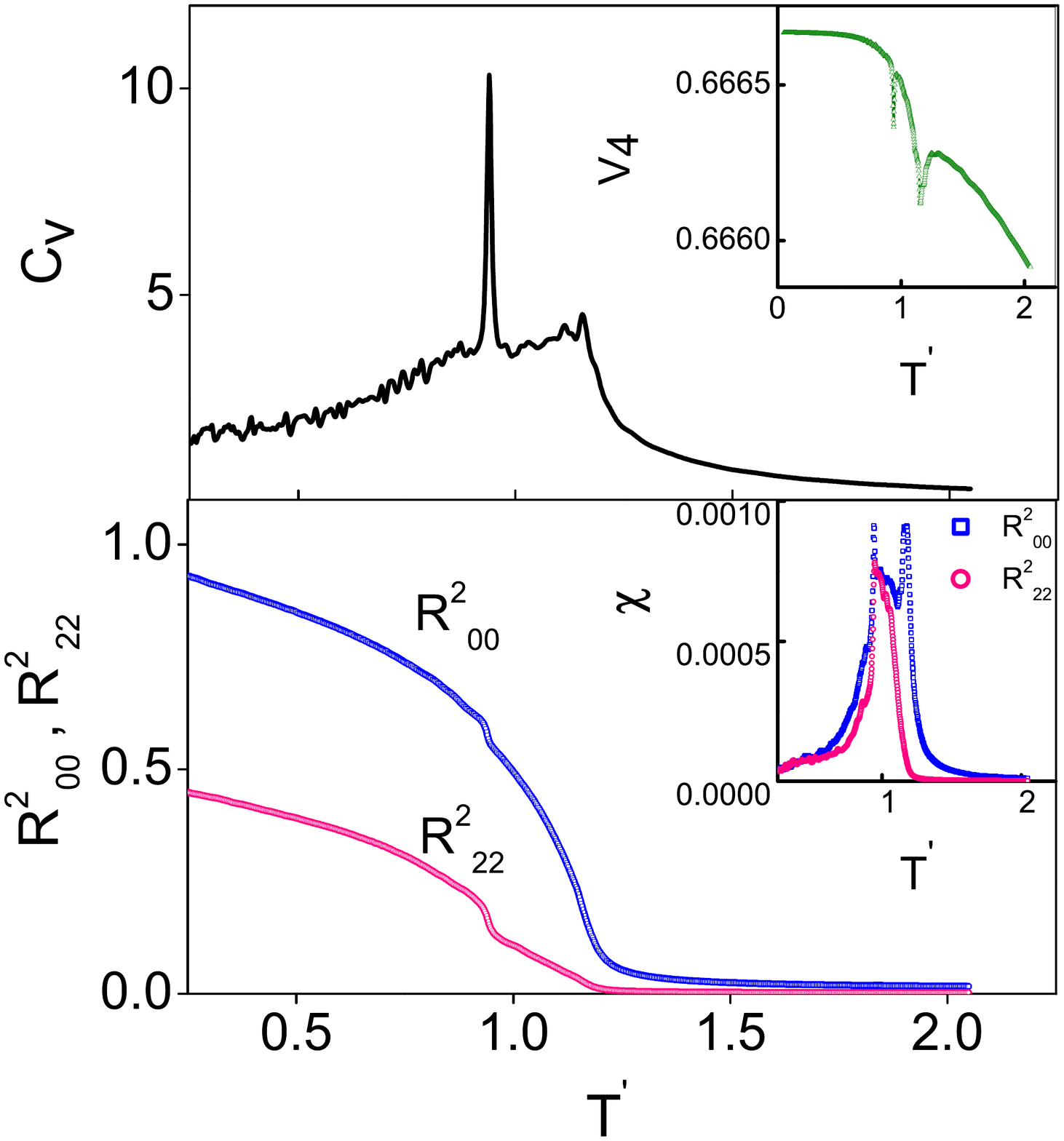} 
\label{fig:7d}}
\subfigure[]{
   \includegraphics[scale=0.22]{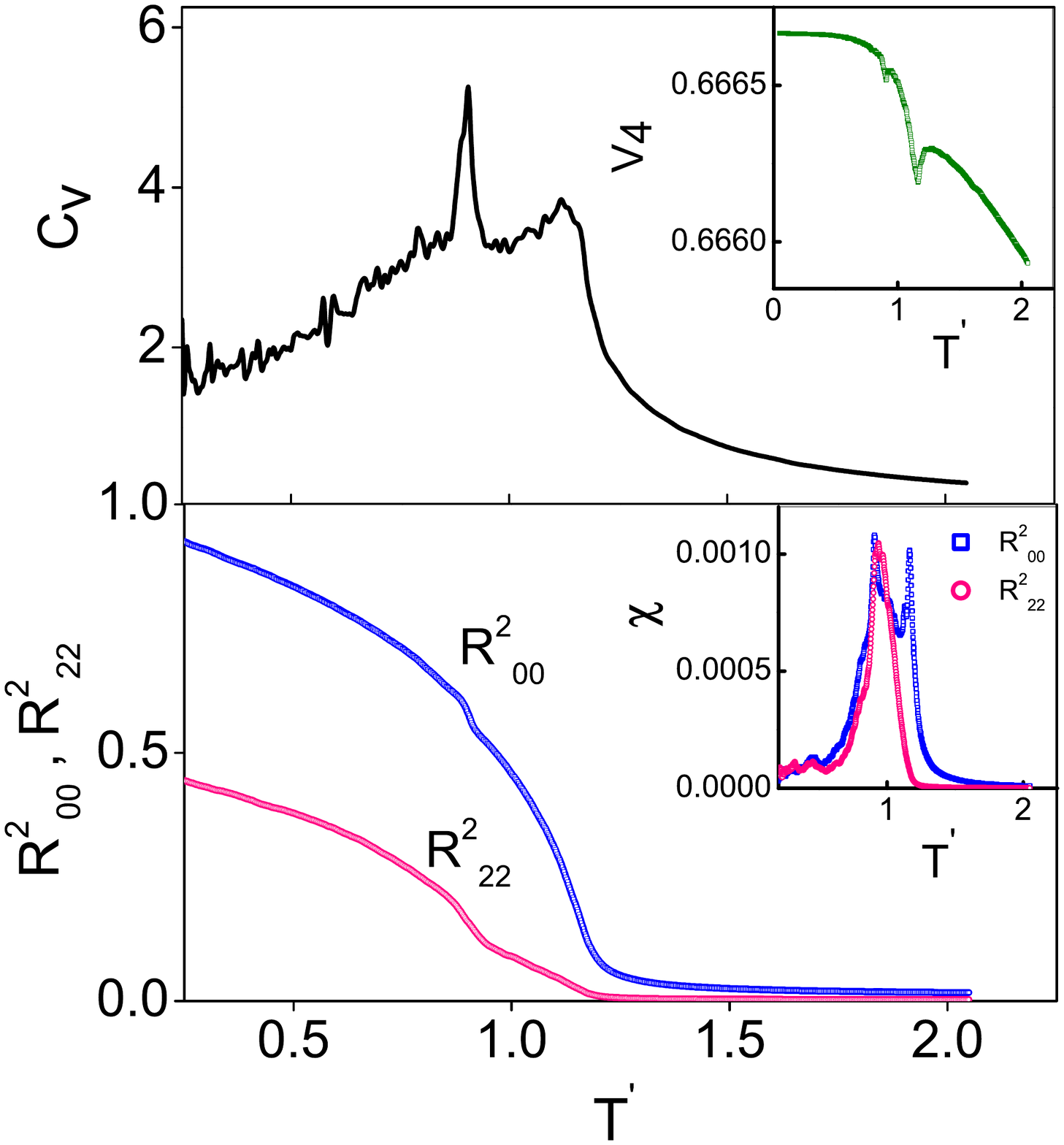} 
\label{fig:7e}}
\subfigure[]{
   \includegraphics[scale=0.22]{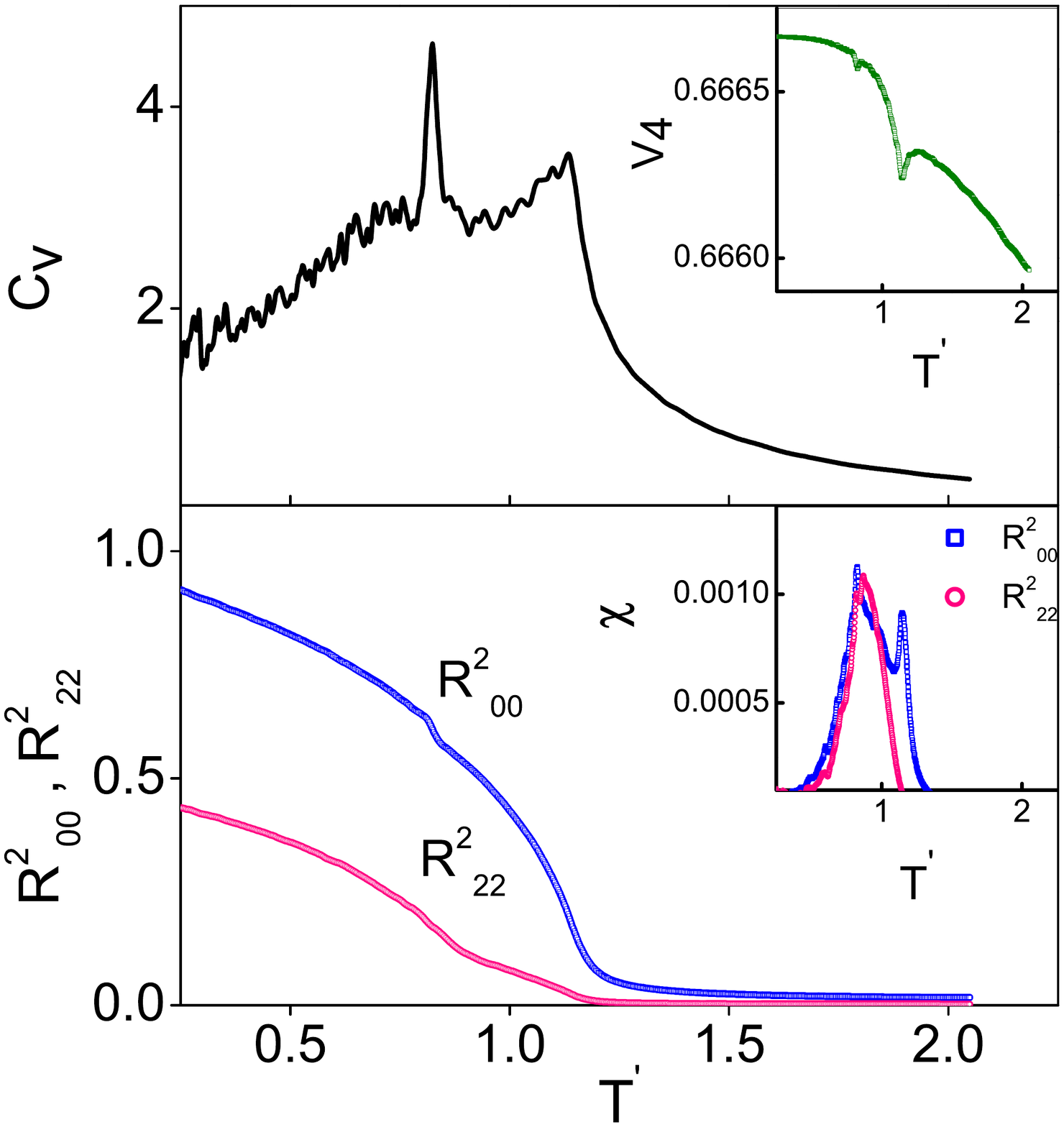} 
\label{fig:7f}}
} 
\caption{ (color online) Specific heat profile with (inset) energy 
cumulant $V_{4}$ and order parameters with (inset) susceptibility profiles 
at different  $\lambda^{*}$ values (a) 0.54; (b) 0.58; (c) 0.62; 
(d) 0.66; (e) 0.69; and (f) 0.72} 
\label{fig:7}
\end{figure*}

     Referring to  Fig.~\ref{fig:2d}, we observe that the $C_{v}$ peak
splits starting from $\lambda^{*} \sim 0.54$, signalling the onset of 
two transitions. The temperature gap between transition peaks increases
with $\lambda^{*}$ above this value, attaining a maximum 
at $\lambda^{*}$ = 0.733 (T on the triangle). These observations are 
illustrated in Fig.~\ref{fig:7}, plotting all the relevant variables 
as a function of temperature at chosen values of $\lambda^{*}$ 
(0.54, 0.58, 0.62, 0.66, 0.69, 0.72)  along the segment $C_{3}$T. These 
graphs depict the temperature variation of specific heat $C_{v}$ with energy 
cumulant $V_{4}$ (inset) and order parameter ($R^{2}_{00}, R^{2}_{22}$) 
profiles along with respective susceptibilities $\chi$ (inset).
      
The  nature of the two phases below the clearing point is inferred from
the order parameter profiles and their susceptibility peaks.  Referring 
to the two transition temperatures in decreasing order as $\textit{T}_{1}$ and 
$\textit{T}_{2}$, the data indicate that the onset of a biaxial phase takes place 
at $\textit{T}_{1}$ itself, and the growth of biaxial order in the intermediate 
phase is marginal as compared to the uniaxial order. Further, both the
uniaxial and biaxial order parameters display a sudden upward jump at 
$\textit{T}_{2}$, and subsequently increase rapidly (more pointedly the biaxial 
order $R^{2}_{22}$) as the temperature is lowered further. This behaviour is
prominent  in the neighbourhood of  $\lambda^{*}$ = 0.66.  The 
susceptibility of $R^{2}_{00}$ exhibits two peaks corresponding to the 
two transitions, whereas that of  $R^{2}_{22}$ shows only a single peak 
at $\textit{T}_{2}$, for all values of $\lambda^{*}$. The energy cumulant
$V_{4}$ shown in the  inset of each of the figures indicates the first 
order nature of the $I - N_{B1}$ transition (at $\textit{T}_{1}$). The 
additional second dip at the lower temperature  transition (at 
$\textit{T}_{2}$) appears to point towards the progression of the first 
order nature of the $N_{B1} - N_{B}$ transition. It is observed that 
the dip in the cumulant at $\textit{T}_{2}$ is maximum at $\lambda^{*}$= 0.66.

\begin{figure}
\centering {\subfigure[]{
\includegraphics[scale=0.3]{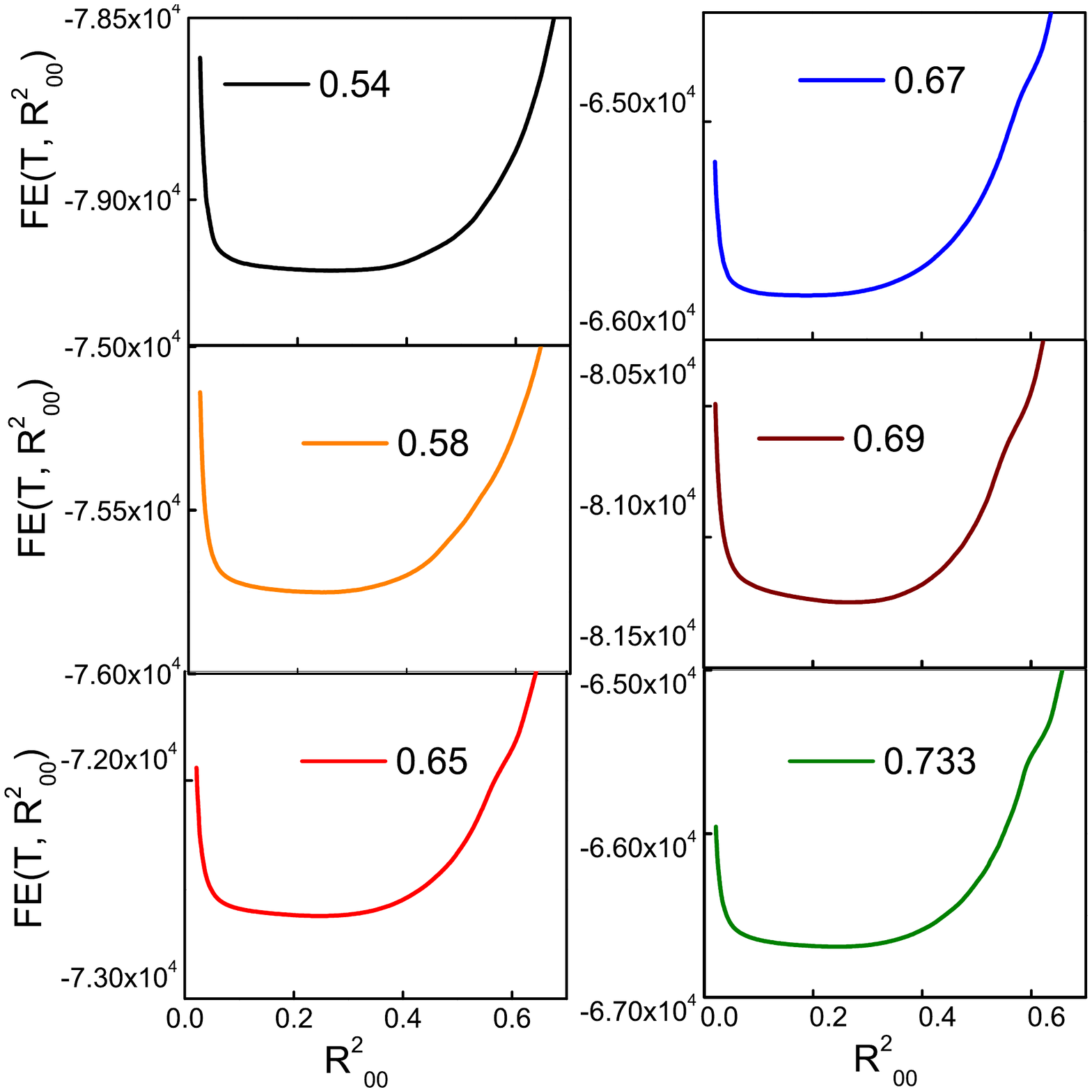}
\label{fig:8a}}
 \subfigure []{
   \includegraphics[scale=0.3]{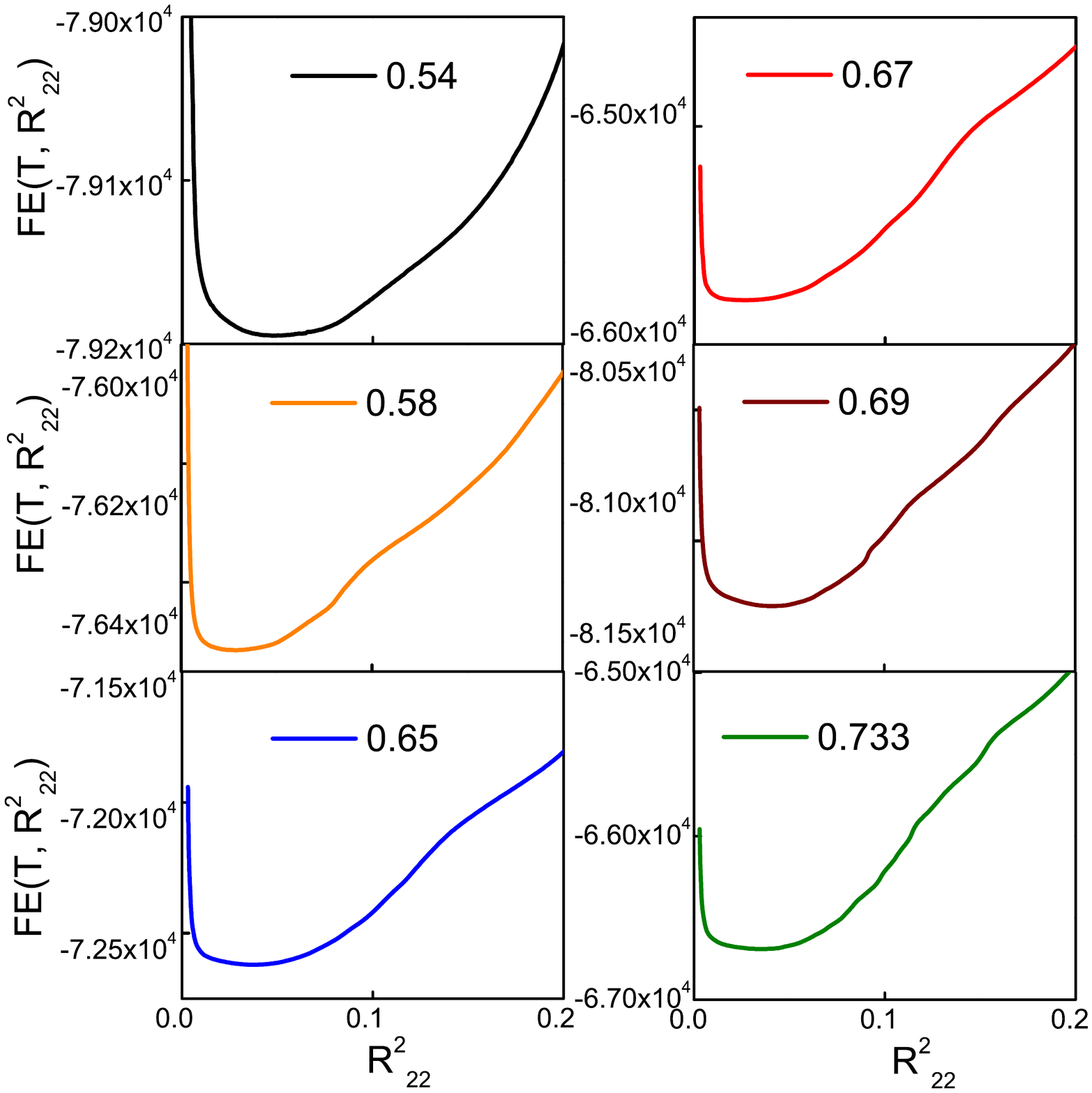}
\label{fig:8b}}
}
\caption{(color online) Free energy shown as a function of 
(a) $R^{2}_{00}$ and (b) $ R^{2}_{22}$, at the transition temperature  
$\textit{T}_{1}$ for $\lambda^{*}$ values in the region $C_{3}$T
of Fig.~\ref{fig:1}. } 
\label{fig:8}
\end{figure}
       
  We also examined the representative free energy plotted as a function of 
 the order parameters at the transition temperatures $\textit{T}_{1}$
 and $\textit{T}_{2}$. These variations observed near $\textit{T}_{1}$, for different 
 $\lambda^{*}$ values (covering the region $C_{3}$T) are shown in 
 Figs.~\ref{fig:8}. Focussing on Figs.~\ref{fig:8a} and \ref{fig:8b}, 
 one immediately observes that both the free energy profiles 
 (with respect to $R^{2}_{00}$ and $R^{2}_{22}$) at $\textit{T}_{1}$ 
 show a distinctive indication of a developing
 minimum at a lower temperature evidenced by the systematic deviations 
 (from a smooth continuation) of the profiles at the respective higher values
 of the two order parameters. And, as $\lambda^{*}$ increases in the 
 $C_{3}$T region, the location of these sharp deviations progressively
 shift to a higher value of the corresponding order parameter. 
 
 \begin{figure}
\centering {\subfigure[]{
\includegraphics[scale=0.3]{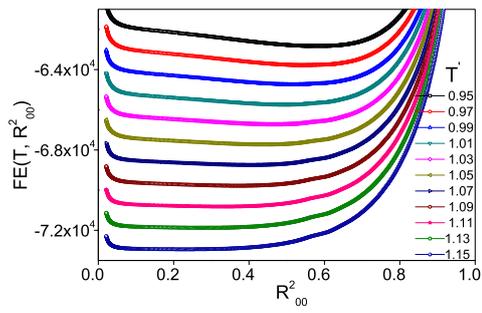}
\label{fig:9a}}
 \subfigure []{
   \includegraphics[scale=0.3]{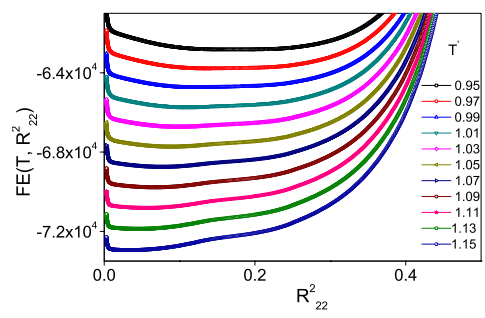}
\label{fig:9b}}
}
\caption{(color online) Free energy shown as a function of 
(a) $R^{2}_{00}$  and (b) $ R^{2}_{22}$, on cooling from 
$\textit{T}_{1}$ to $\textit{T}_{2}$ for $\lambda^{*}$ = 0.65}
\label{fig:9}
\end{figure}   
 
 \begin{figure}
\centering {\subfigure[]{
\includegraphics[scale=0.3]{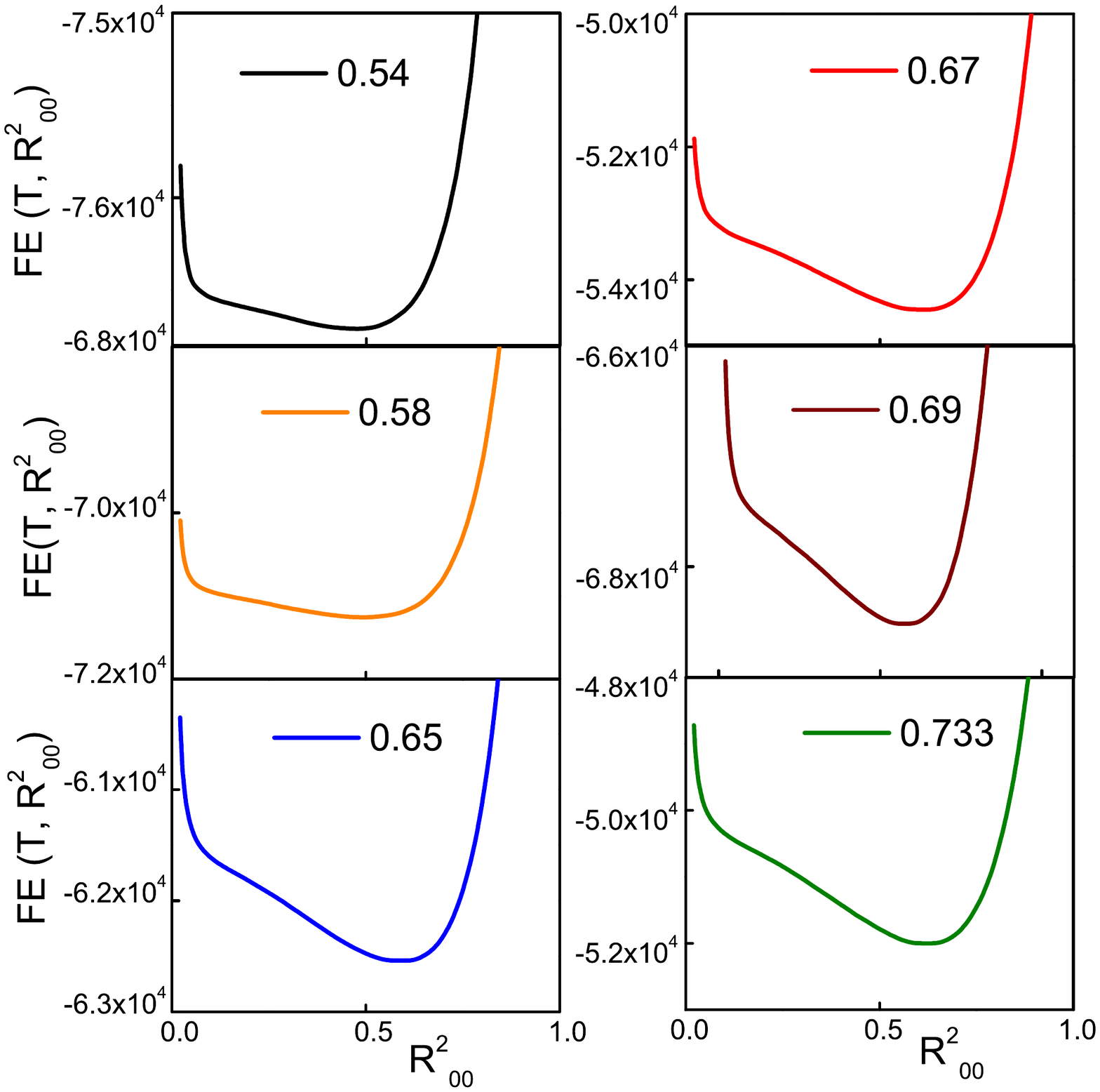}
\label{fig:10a}}
 \subfigure []{
   \includegraphics[scale=0.3]{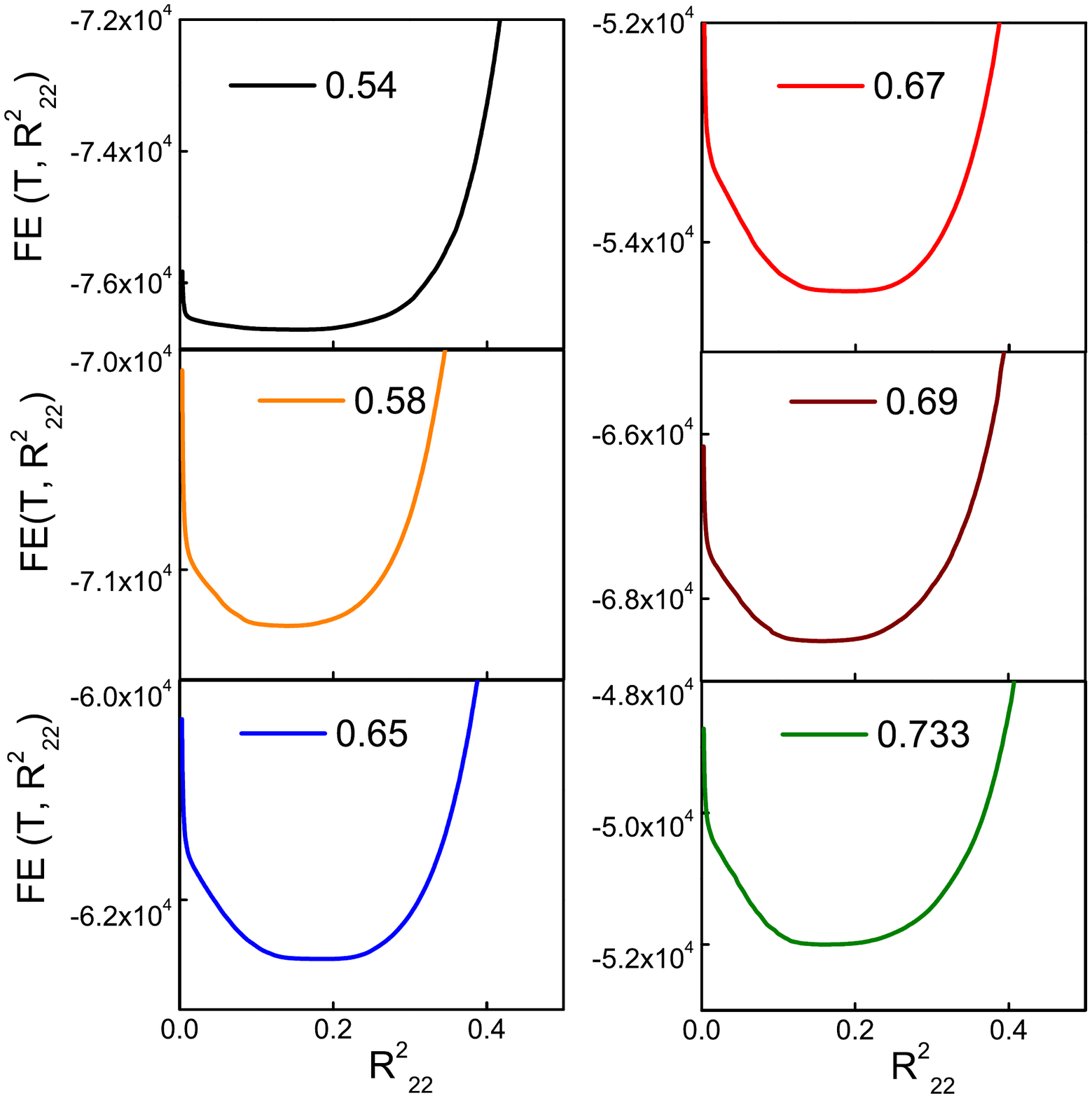}
\label{fig:10b}}
}
\caption{(color online) Free energy shown as a function of 
(a) $R^{2}_{00}$ and (b) $ R^{2}_{22}$, at the transition temperature  
$\textit{T}_{2}$ for $\lambda^{*}$ values in the region $C_{3}$T of 
Fig.~\ref{fig:1}. } 
\label{fig:10}
\end{figure}
       
 We tracked 
 the variation of these profiles closely from $\textit{T}_{1}$ to $\textit{T}_{2}$ 
 (shown in Fig.~\ref{fig:9} for a single value of $\lambda^{*}$=0.65), and 
 find that the free energy minima gradually shift 
 towards high order regions, and the second transition at $\textit{T}_{2}$ 
 corresponds to a gradual shift of the free energy minima towards the 
 curious regions, depicted in Fig.~\ref{fig:8}. The variation of the free 
 energy at $\textit{T}_{2}$ is shown  in Figs.~\ref{fig:10a} and \ref{fig:10b}
 for different values of  $\lambda^{*}$ (in region $C_{3}$T). These depict
 a free energy minimum attained at $\textit{T}_{2}$ for all values of $\lambda^{*}$.
    
  We argue that the progression of the free energy profiles with temperature, 
  as a function of $R^{2}_{00}$ and ${R}^{2}_{22}$, and matching of the values of 
 respective order parameters at $\textit{T}_{2}$ with the location of  sharp deviations observed
 in Fig.~\ref{fig:8} are further evidences for the existence of two 
 transitions in this region. It may be pointed this could be made 
 possible only by adopting a MC sampling procedure which facilitates 
 the computation of free energy profiles of the system, $\textit{via}$
 the density of states.
 
      We are thus led to the conclusion that in this region 
 of $\lambda^{*}$ values, the medium undergoes two transitions, and both 
 the low temperature phases have biaxial symmetry. From the data on the
 limited temperature region available for the intermediate phase, and 
 in comparison with the low temperature phase, it appears that the
 biaxial order in the intermediate phase is somewhat inhibited, presumably 
 by free energy barriers. It is only after the second transition at $\textit{T}_{2}$
 (between the two biaxial phases, and hence necessarily a first order transition)
 that the biaxial order shows normal increase with decrease in 
 temperature, as is to be expected. Thus we propose the phase sequence 
 in this region to be $N_{B}-N_{B1}-I$.

 \begin{figure}
\centering
\includegraphics[scale=0.5]{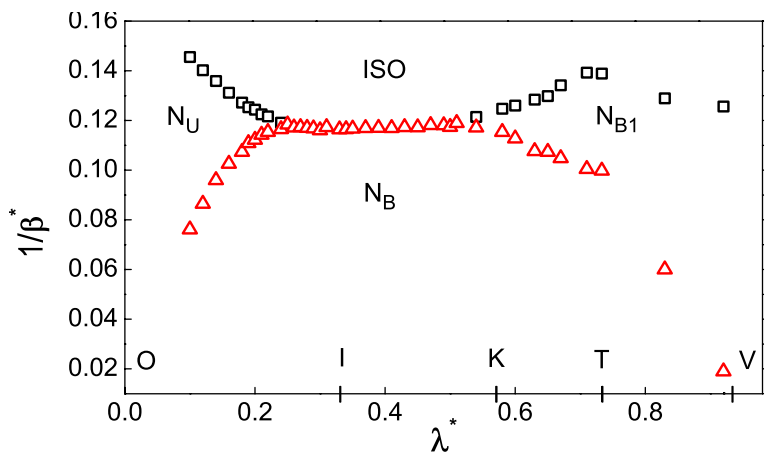}
\caption{(color online) Phase diagram  as a function of $\lambda^{*}$, 
derived from RW-ensembles. The transition temperature ${1}/{\beta^{*}}$ 
is scaled to conform to mean - field values as indicated in the  text. 
Points along OIV in Fig.~\ref{fig:1} are mapped onto the $\lambda^{*}$-axis 
for reference. An additional biaxial-biaxial transition is observed 
in the region KTV in place of a single transition (to the biaxial phase) 
predicted by the mean-field theory \cite{kamala14}.}
\label{fig:11}
\end{figure} 
 
We now construct the phase diagram as a function of the arc length $\lambda^{*}$ 
based on the specific heat data (from RW-ensembles), shown in  
Fig.~\ref{fig:11} at
56 values of $\lambda^{*}$ distributed over the arc OIV (see \cite{kamala14} 
for details).  The transition temperatures  at a few representative
values of $\lambda^{*}$ beyond the Landau point T (segment TV) are 
obtained from the B-ensemble data. The
temperature  $\textit{T}^{'}$ of the simulation is  scaled to conform to  
the values ${1}/{\beta^{*}}$ used in the mean field  treatment as 
discussed in section III. 

    A comparison of the phase diagram proposed from the current MC 
simulations \cite{kamala14} with the one predicted based on mean-field 
theory \cite{matteis07} brings out clear qualitative differences in the 
region $C_{3}$TV of the  essential triangle. We observe that the predicted 
direct transition from the isotropic to biaxial phase is replaced by 
two transitions in  which an intermediate biaxial phase occurs between 
these two phases. These results start deviating starting from 
$\lambda^{*} \gtrsim 0.54$, very close to the point $C_{3}$ in Fig.~\ref{fig:1}. 
The fact that B-ensembles constructed from simple configurational random walks 
based on Metropolis algorithm fail to detect the second transition
 in this $\lambda^{*}$ region  merits some discussion.

\section{ Discussion}

\begin{figure}
\centering{
\includegraphics[scale=0.3]{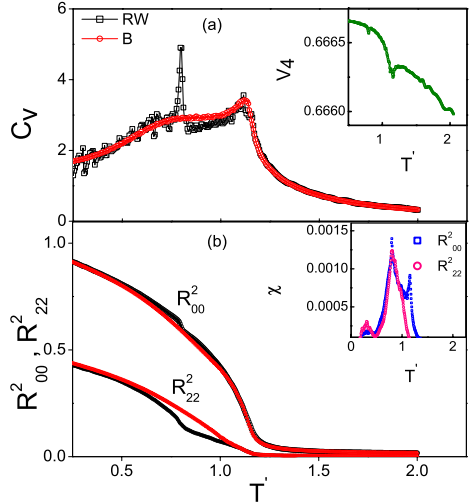}
}
\caption{(color online) Comparison of data, as a function of temperature, 
from the B- and RW- ensembles, at the Landau point T ($1/3, 1/9$): 
(a) Specific heat $C_{v}$  and (b) order parameters $R^{2}_{00}$ and 
$R^{2}_{22}$. The insets focus on (a) the energy cumulant $V_{4}$ and 
(b) order parameter susceptibilities ($\chi '$s), both derived from 
RW-ensembles ($\lambda^{*}$ = 0.733).}
\label{fig:12}
\end{figure}
  
In order to look for  the origin of the additional low temperature 
specific heat peak observed in the region $C_{3}$T, which was not 
detected by Boltzmann sampling, we made a 
comparison of the simulation results from RW-ensembles 
with those obtained from B-ensembles at $\lambda^{*}$=0.733 $(1/3,1/9)$ 
(Landau point T) shown in Fig.~\ref{fig:12}. The location of T is unique 
as it is the intersection point of the dispersion parabola
with the segment IV. MF theory predicts a direct transition from the isotropic
to biaxial phase at this point, and it is the only such point on the parabola. From the perspective of the 
interaction Hamiltonian, the coordinates $(\gamma, \lambda)$ of T 
represent a unique symmetry: The Hamiltonian has no interaction between the 
uniaxial components of the molecular tensor ($\mu = 0$ at T, see 
Eqn.~\eqref{eqn:w5}), and is purely biaxial in nature (involving $\bm{m}$ 
and $\bm{e_}\perp$ axes). A consequence of the present findings in this 
context is curious. They do confirm the presence of a direct transition 
from the isotropic to biaxial symmetry, but these also indicate that 
there is yet another biaxial-to-biaxial transition at a lower temperature. 
Further, the onset of the first biaxial phase at $\textit{T}_{1}$ is not leading 
to a natural progression of the biaxial order with decrease in temperature, 
and it is only after the transition at $\textit{T}_{2}$ that the 
macroscopically significant and hence observable, $R^{2}_{22}$ value 
seems to be realizable.
     
     We thus focus on the Landau point, and present the simulation results
obtained from the two types of MC sampling methods: data from 
B-ensembles and from RW-ensembles.  Fig.~\ref{fig:12} shows the  
specific heat (energy cumulant as inset) and order parameters
(susceptibilities as inset), computed as a function of temperature at 
the point T, obtained from these ensembles. It may be noted that the derived
physical variables from B-ensembles do not betray the onset of the second
transition at $T_{2}$, thus lending support to    MF predictions, as has 
been noted in the earlier report on this  work \cite{kamala14}. 
 
 We now examine the contour maps of the distribution of microstates
in the entropic ensemble (set of microstates which are approximately 
uniformly distributed with respect to energy) collected at the Landau 
point.
\begin{figure}
\centering{
\includegraphics[scale=0.6]{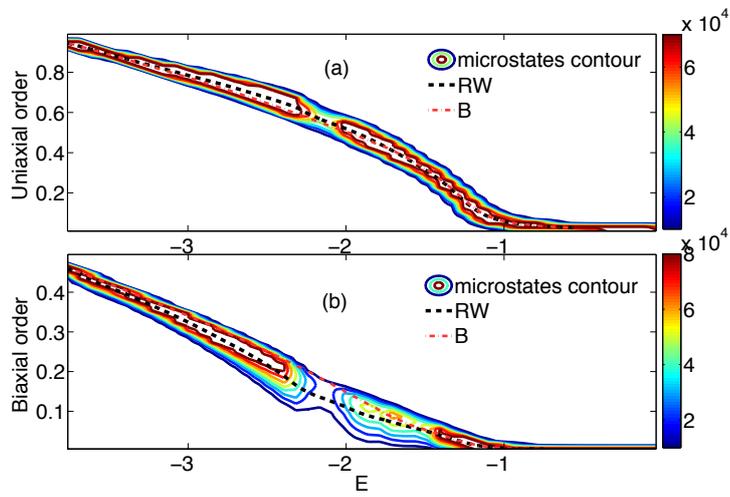}
}
\caption{(color online) Contour plots of the distribution of microstates  
collected in the entropic ensemble at  $\lambda^{*}\simeq0.733$: 
(a) Microstate energy versus its uniaxial order and (b) Microstate 
energy versus its biaxial order. The superimposed red (dash dotted line)
and black (dashed) lines are thermal averages from B- and RW-ensembles, 
respectively.} 
\label{fig:13}
\end{figure}
Fig.~\ref{fig:13}(a) depicts such a contour map  in the space of
uniaxial order  and energy (per site), along with the thermal averages  
computed from RW-ensembles and B-ensembles superposed for ready comparison. 
Similarly, Fig.~\ref{fig:13}(b) shows corresponding contour map plotted 
between biaxaial order and energy (per site), along with thermal averages 
of the two canonical ensembles again superposed.  The traversal path of 
the B-ensemble averages is seen to be encompassing regions corresponding 
to contour peak positions, whereas the RW-ensemble average is observed 
to follow a different trajectory, consequent to encompassing a wider 
collection of microstates visiting sparse regions, corresponding to 
large deviations of the order parameter. This is seen as a manifestation 
of the process of collection of microstates of the entropic ensemble
by the algorithm employed, representative in their distribution 
(with respect to energy) of the underlying density of states. As has 
been pointed out and argued earlier (Fig.~6 in \cite{kamala14}), 
the algorithmic guidance of the WL-procedure is seeking out all accessible 
microstates (an approximate microcanonical ensemble) in each bin of 
energies, in the process visiting relatively rare states which correspond 
to larger excursions in the order
parameter, and hence correspondingly larger fluctuations of the component 
energies of the Hamiltonian in Eqn.~\eqref{eqn:w6}, while conforming to the 
same energy bin.  The requirement of an accurate determination of DoS through 
the entropic sampling procedure apparently demands inclusion of these rare 
microstates, and the process of reweighting used to construct the equilibrium 
ensembles through this elaborate procedure, includes them in the thermal 
averages as a consequence. 
 
        Thus the differences observed in the averages from 
the two procedures are to be appreciated from the standpoint of simulations.
As has been discussed \cite{kamala14}, these rare microstates 
indeed correspond to situations where the ordering  of either
of the molecular axes (involved in the $D_{4h}$ symmetry of the pair-wise
interaction, i.e $\bm{m}$ or $\bm{e}_{\perp}$ ) form a spontaneous and
equally probable calamitic axis during the evolution of the system, by 
virtue of having the largest instantaneous  eigenvalue of the diagonalised
ordering tensors of the three molecular axes. It is apparent that 
conventional sampling methods, not under algorithmic compulsion to estimate 
the DoS of the system, are not geared to sample
such rare states, and hence come up with different averages.
In the process, it appears that second low temperature transition 
is not evident in the earlier work.

     In addition, we note that the deviation of the simulation results from
the mean field expectations occur along the diagonal IV of the essential
triangle where the pairwise interaction Hamiltonian has $D_{4h}$ 
symmetry and is  expressed in the reduced form as in Eqn.~\eqref{eqn:w5} 
in terms of a single
parameter $\mu$. It is observed that the deviations start from 
$\lambda^{*}\geq 0.54$  (point $C_{3} \ (5/29, 19/87)$ and 
continue till the Landau point where $\mu = 0$. It may be noted that the 
point K $(0.2, 0.2)$ corresponding to  $\mu = -1$ is very  close to the 
point $C_{3} \ (0.172, 0.218)$. It can be inferred from 
Eqn.~\eqref{eqn:w7} that, starting from the neighbourhood of K, the uniaxial attractive
coupling of  the $\bm{e}$-axis  becomes lower in strength than that of 
the (biaxial) attractive coupling of the other two axes, and continues to 
decrease as $\lambda^{*}$ increases  on the diagonal. As a result, the 
ordering of the biaxially coupled $\bm{e}_{\perp}$  and  $\bm{m}$ axes 
is favoured as temperature is decreased, leading to the first onset of a 
biaxial symmetry from the isotropic phase. This is followed by an 
ordering of the $\bm{e}$-axes at a lower temperature, leading to the 
stabilization of both the orders, in particular biaxial order. Thus it 
appears that the growth of biaxial order in the intermediate biaxial phase is 
inhibited by the lack of long range order of the $\bm{e}$-axes (in this 
region of $\lambda^{*}$-axis). As one approaches the Landau point, 
$\mu$ in Eqn.~\eqref{eqn:w5} tends to zero (from the negative side), thereby 
suppressing the  second transition temperature as well as weakening the 
efficacy of this term to drive a transition. This lack of concomitant 
ordering of all the molecular axes leads to inhomogeneity in the medium,
and we tend to attribute all the interesting aspects of the simulation to this 
feature of the Hamiltonian.

            Finally, we wish to comment on the curious role played
 by the WL-algorithm in the analysis of the phase diagram. It has been 
 already established  that this algorithm assists the system in overcoming 
 energy barriers of the system, as the simulation pushes the system to 
 make a random walk in the configuration space which is uniform with respect to 
 energy. A successful convergence of the probability density yields a 
 limiting distribution of microstates with respect to the total energy of 
 the system - the representative density of states. The role of this algorithm 
 in the present study seems to be qualitatively different and yet illustrative 
 of its varied applicability. The WL algorithm, even while operating within a 
 single energy bin (an approximate microcanonical ensemble), appears to seek out 
 rare states, corresponding to the otherwise inaccessible fluctuations of the 
 component energies, making up the total energy (see Fig.~ 6 in
  \cite{kamala14}). Inclusion of these microstates ( as representative states 
  for purposes of computing averages) is naturally embedded in the
  WL-method, while estimating the DoS accurately. We argue that the 
  Metropolis sampling  fails to access these states due to apparent energy 
  barriers within the system inhibiting sampling of microstates with 
  such large fluctuations in their energy components. We conclude that the 
  new results reported here are the outcome of this facet of 
  efficiency of the entropic sampling.
           
   \section{Conclusions} 
In conclusion, we present compelling  evidences from Monte Carlo simulations 
based on entropic sampling, to propose an additional biaxial phase  along 
a region of the arc of the essential triangle,  augmenting our earlier 
report \cite{kamala14}. The arguments advanced in this respect, particularly 
of the inevitable presence of inhomogeneities in the absence of a long-range 
order of the third stabilising axis $\bm{e}$, seem to lend support to the 
findings (based on Boltzmann MC sampling) reported in the partly repulsive 
region of the $\lambda^{*}$-axis (segment TV: the $\mu$-model \cite{Dematteis08}). 
At a more general level we conclude that the cross-coupling between the 
uniaxial and biaxial tensor components of the neighbouring molecules ($\gamma$-term in Eqn.~\eqref{eqn:w6}) seems to be playing an important 
role in determining the phase sequences. Further, we suggest that its 
significant presence, even along trajectories inside the triangle 
(which could be relevant for practical purposes) should have such inhibitive 
influence on the condensation of a biaxial phase with measurable biaxial 
order. Our recent simulational work on two such trajectories interior 
to the triangle are supportive of this conjuncture (to be published).
         
 \section {acknowledgments}
We thank Professor N. V. Madhusudana (Raman Research Institute, Bangalore, India)
for useful discussions. The simulations are carried out in the Centre for
Modelling Simulation and Design, University of Hyderabad.

\end{document}